\documentclass[letterpaper, 10 pt, conference]{ieeeconf}  % Comment this line out if you need a4paper

\IEEEoverridecommandlockouts                              % This command is only needed if 
\usepackage{graphics} % for pdf, bitmapped graphics files
\usepackage{epsfig} % for postscript graphics files
\usepackage{caption}
\usepackage{subcaption}
\usepackage{textcomp}
\usepackage{xcolor}       % No options
\usepackage{colortbl}     % Enables \rowcolor, \cellcolor, etc.
\definecolor{myorange}{rgb}{1.0, 0.5, 0.0} % Define a custom orange
\usepackage{hyperref}
\hypersetup{
    colorlinks=true,
    linkcolor=blue,
    filecolor=blue,
    urlcolor=blue,
    citecolor=blue
}

\usepackage{times} % assumes new font selection scheme installed
\usepackage{bm}
\usepackage{cite} 
\usepackage{amsmath} % assumes amsmath package installed
\usepackage{amssymb}  % assumes amsmath package installed
\usepackage{mathrsfs} 
\usepackage{dsfont}
\usepackage{stfloats}
\usepackage{comment}
\newcommand{\bfF}{{\bf{F}}}
\newcommand{\bfv}{{\bf{v}}}
\newcommand{\bfA}{{\bf{A}}}
\newcommand{\bfB}{{\bf{B}}}
\newcommand{\bfT}{{\bf{T}}}
\newcommand{\bfY}{{\bf{Y}}}
\newcommand{\bfV}{{\bf{V}}}
\newcommand{\bfy}{{\bf{y}}}

\newcommand{\bfD}{{\bf{D}}}
\newcommand{\bfR}{{\bf{R}}}
\newcommand{\bfb}{{\bf{b}}}
\newcommand{\bmeta}{{\bm{\eta}}}

\newcommand{\bfx}{{\bf{x}}}
\newcommand{\bfg}{{\bf{g}}}
\newcommand{\der}{{\mathrm{d}}}

\title{\LARGE \bf
 Sparsity-Promoting Dynamic Mode Decomposition Applied to Sea Surface Temperature Fields}

\author{Zhicheng Zhang, Yoshihiko Susuki, and Atsushi Okazaki
\thanks{This~work~was supported by~JST Moonshot~R\!\&\!D,~Grant~No.~JPMJMS224.}% <-this % stops a space
\thanks{Z.~Zhang and Y.~Susuki are with
the Department of Electrical~Engineering, 
Kyoto University, Katsura, Nishikyo-ku, Kyoto 615-8510, Japan~(e-mails:~\texttt{zhang.zhicheng.2c@kyoto-u.ac.jp; susuki.yoshihiko.5c@kyoto-u.ac.jp}).}%
\thanks{A. Okazaki is with Institute of Advanced Academic Research, Center for Environmental Remote Sensing, Chiba University, 1-33 Yayoi-cho, Inage, Chiba 263-0022, Japan
(e-mail: \texttt{atsushi.okazaki@chiba-u.jp}).}
}

\begin{document}

\maketitle
%\thispagestyle{empty}
%\pagestyle{empty}

%\maketitle

%%%%%%%%%%%%%%%%%%%%%%%%%%%%%%%%%%%%%%%%%%%%%%%%%%%%%%%%%%%%%%%%%%%%%%%%%%%%%%%%
 \begin{abstract}
In this paper, we leverage Koopman mode decomposition to analyze the nonlinear and high-dimensional climate systems acting on the observed data space. The dynamics of atmospheric systems are assumed to be equation-free, with the linear evolution of observables derived from measured historical long-term time-series data snapshots, such as monthly sea surface temperature records, to construct a purely data-driven climate dynamics. 
In particular, sparsity-promoting dynamic mode decomposition is exploited to extract the dominant spatial and temporal modes, which are among the most significant coherent structures underlying climate variability, enabling a more efficient, interpretable, and low-dimensional representation of the system dynamics.
We hope that the combined use of Koopman modes and sparsity-promoting techniques will provide insights into the significant climate modes, enabling reduced-order modeling of the climate system and offering a potential framework for predicting and controlling weather and climate variability.
 \end{abstract}

%\begin{IEEEkeywords}

%%%%%%%%%%%%%%%%%%%%%%%%%%%%%%%%%%%%%%%%%%%%%%%%%%%%%%%%%%%%%%%%%%%%%%%%%%
\section{Introduction}
Discovering dominant coherent structures with dynamical and physical implications in climate systems, 
such as intraseasonal and Madden Julian Oscillation, seasonal El Ni\~{n}o and La Ni\~{n}a–Southern Oscillation, 
and sea surface temperature anomalies \cite{marshall1961atmosphere}, 
is an important issue in Earth's system modeling, estimation, variability, forecasting, and control \cite{ghil2008climate}, 
and remains a primary focus in atmospheric, ocean, and climate science. 

However, the dynamics of climate's states evolving in a finite dimensional phase space naturally contain nonlinear trends, 
including multiple equilibria, bifurcations, strange attractors, turbulence, and riddled basins.
On the one hand, these complexities inherently make the limitations of traditional methods, such as the linear inverse model (LIM) \cite{tu2014dynamic}, 
in extracting coherent structures and predicting atmospheric processes, especially when it comes to revealing nonlinear behavior.
On the other hand, while practical methods such as those based on empirical orthogonal functions (EOFs) \cite{schmidt2019spectral} 
can provide valuable insights into the dominant modes of variability in climate systems, they often lack a direct connection to the underlying dynamics.

In recent decades, the operator-theoretic approach, with a focus on the Koopman operator \cite{koopman1931hamiltonian}, has gained great attention for analyzing dynamical systems by lifting nonlinear dynamics into a linear representation. This state-of-the-art method describes the linear evolution of observables in an infinite-dimensional functional space \cite{mezic2005spectral,mauroy2020koopman}.
Needless to say, applied Koopmanism played a growing role in uncovering structures in (geophysical) fluid dynamics \cite{rowley2009spectral, mezic2013analysis} and in interpreting coherent modes in climate and weather systems. When Koopman mode decomposition (KMD) \cite{susuki2011nonlinear} is combined with its numerical counterparts, such as various dynamic mode decomposition (DMD) \cite{kutz2016dynamic,jovanovic2014sparsity}, %and kernel-based extensions \cite{williams2016kernel}, 
it provides spectral analysis for decomposing spatiotemporal patterns. Related literature has validated these methods for capturing climate-related phenomena, including the Madden–Julian Oscillation (MJO) \cite{lintner2023identification} and its dominant component, tropical intraseasonal variability \cite{alexander2017kernel, giannakis2015spatiotemporal}, the El Niño–Southern Oscillation (ENSO) \cite{wang2020extended,froyland2021spectral}, sea surface temperature (SST) anomalies \cite{navarra2021estimation}, the Pacific Decadal Oscillation (PDO) \cite{navarra2024variability}, and sea ice cover variability \cite{hogg2020exponentially}.
Notably, Navarra et al.~\cite{navarra2021estimation} estimated the Perron-Frobenius and Koopman operators based on one-dimensional Ni\~{n}o-3 time series data and monthly mean Pacific SST, where the system evolution density was derived from the transfer or Koopman eigenfunctions. Subsequently, the variability of tropical ocean and PDO of global SST were identified via Koopman modes \cite{navarra2024variability}, and extended to examine the continuous spectrum \cite{mezic2005spectral,colbrook2024rigorous}.

Building on these results, this paper focuses on the performance and model reduction of KMD applied to real-world, data-driven climate dynamics.
More precisely, we examine three distinct time scales associated with different observable spaces: the variability of SST fields on intraseasonal or monthly (high resolution), seasonal (medium resolution), and annual (low resolution) cycles.
Notably, our work differs from existing SST studies using kernel methods \cite{navarra2021estimation, navarra2024variability} by taking sparsity-promoting dynamic mode decomposition (SPDMD) \cite{jovanovic2014sparsity} method to extract leading Koopman modes and illustrate climate variability.  
This is accomplished by penalizing an additional $\ell_1$-norm in the least-squares error minimization, which promotes sparsity and ensures the selection of the most significant modes with \emph{nonzero amplitudes}, rather than relying on the {magnitude of eigenvalues} as in standard DMD \cite{kutz2016dynamic}.
One of the benefits of SPDMD is that it can {make a trade-off between accuracy (i.e., performance loss with respect to residual error) and model complexity (i.e., the number of modes)} by tuning the sparsity weights, offering users flexible choices.
Moreover, by comparing methods such as DMD and companion-based DMD (CDMD) \cite{rowley2009spectral}, we demonstrate that SPDMD provides an effective approach for identifying dominant Koopman modes and deriving a low-dimensional climate dynamics from SST data fields.

The rest of this paper is organized as follows: 
Section~\ref{sec:methods-DMD-SPDMD} introduces the DMD method for approximating the KMD, and deriving a low-dimensional linear system via SPDMD.
%Section~\ref{sec:methods-DMD-SPDMD} introduce the methods with respect to DMD and SPDMD to approximaite the KMD to obtain Koopman modes and the concepts of KMD and SPDMD for model reduction to derive low-dimensional linear systems.
%Section~\ref{sec:sparsity-promoting-model-reduction} introduces SPDMD for model reduction to derive low-dimensional linear systems. 
Section \ref{sec:SST_applications} presents three experiments on SST data, extracting dominant Koopman modes to analyze spatial and temporal structures in climate systems. 
Section \ref{sec:conclusion} concludes the paper.
Finally, some preliminaries on the Koopman operator are provided in the Appendix for interested readers.

\section{Methodologies}\label{sec:methods-DMD-SPDMD}
\subsection{Dynamic Mode Decomposition}\label{appendix:DMD}
\begin{comment}
   Following the procedures by Klus \emph{et al.} \cite{klus2016numerical}, it is available to collect the discrete-time snapshots of systems \eqref{eq:DT-systems}, i.e., a {finite} dataset $\{\bfx_{k},\bfT(\bfx_{k})\}_{k=0}^{N-1}$ such that  
\end{comment}
We start by collecting discrete-time snapshots, that is, a finite sequence of data pairs $\{\bfx_k, \bfT(\bfx_k)\}_{k=0}^{N-1}$ satisfying $\bfy_k = \bfx_{k+1} = \bfT(\bfx_k)$, which can be concisely represented using the following matrices:
\begin{align}
\begin{aligned}
    \bfY&=
    \begin{bmatrix}
        \bfy_{0}& \bfy_{1}& \ldots& \bfy_{N-1}
    \end{bmatrix},\\
    \bfY^{+}& =\begin{bmatrix}
        \bfy_{1}& \bfy_{1}& \ldots & \bfy_{N}
    \end{bmatrix},
    \label{eq:data-snapshots}
    \end{aligned}
\end{align}
where $\bfY$,$\bfY^{+}\in{\mathbb{R}}^{p\times{N}}$ and typically $p\gg{N}$.
The concept is to interpret the observed data as sampled snapshots of the studied system’s process, governed by an {unknown} dynamics $\bfT(\cdot)$. %or $\bfF(\cdot)$. 
From \eqref{eq:data-snapshots}, the resultant evolution can be derived by a discrete-time {linear time-invariant} system 
% $\bfy_{k+1}=A\bfy_{k}$, resulting in a compact expression 
\begin{align}
 \bfy_{k+1}=\bfA\bfy_{k},
 \quad \Longrightarrow\quad    \bfY^{+}=\bfA\bfY.
    \label{eq:LTI-DMD}
\end{align}
Here $\bfA\in{\mathbb{R}}^{p\times{p}}$ is a finite-dimensional linear approximation of
the Koopman operator $\mathcal{K}$, determined by solving the following \emph{least squares error minimization problem} 
\begin{align}
   \min_{\bfA\in{\mathbb{R}}^{p\times{p}}}\|\bfY^{+}-\bfA\bfY\|_{\text{F}}^{2}.
    \label{eq:least-square-prob}
\end{align}
An exact  solution to problem \eqref{eq:least-square-prob} is estimated by $\bfA= \bfY^{+}\bfY^{\dag}\in{\mathbb{R}}^{p\times{p}}$,
where the symbol ``$\dag$'' is the Moore-Penrose pseudoinverse. Generally, we first project onto a low-dimensional subspace to reconstruct the dominant nonzero eigenvalues and eigenvectors of the matrix $\bfA$.

The numerical method for such exact DMD proposed by Tu \emph{et al.} \cite{tu2014dynamic}, utilizes the singular value decomposition (SVD) with a truncated rank $r$ such that 
$\bfY=\textbf{U}{\bf\Sigma}{\textbf{V}}^{\ast}$, where $\textbf{U}\in{\mathbb{R}}^{p\times{r}}$, ${\bf\Sigma}\in{\mathbb{R}}^{r\times{r}}$, $\textbf{V}\in{\mathbb{R}}^{N\times{r}}$. 
The columns of $\textbf{U}$ and $\textbf{V}$ are orthonormal and ${\bf\Sigma}$ is diagonal.
We obtain $\bfA=\bfY\textbf{V}{\bf\Sigma}^{-1}\textbf{U}^{\ast}$. 
By computing the compression $\tilde{\bfA}=\textbf{U}^{\ast}\bfY\textbf{V}{\bf\Sigma}^{-1}\in{\mathbb{R}}^{r\times{r}}$, it yields the eigen-decomposition $\tilde{\bfA}\textbf{W}=\textbf{W}{\bf\Lambda}$.
Specifically, the linear model \eqref{eq:LTI-DMD} of the dynamics can be reconstructed onto projected coordinates $\tilde{\bfy}=\textbf{U}^{\ast}\bfy$, namely, $\tilde{\bfy}_{k+1}\approx\tilde{\bfA}\tilde{\bfy}_{k}$.
Finally, we have $\bfA={\bf\Phi\Lambda} {\bf\Phi}^{-1}$ and thus
\begin{align}
    \bfA(\bfY\textbf{V}{\bf\Sigma}^{-1}\textbf{W})
    & =\bfY\textbf{V}{\bf\Sigma}^{-1}(\textbf{U}^{\ast}\bfY\textbf{V}{\bf\Sigma}^{-1})\textbf{W} 
    \notag\\
    &=(\bfY\textbf{V}{\bf\Sigma}^{-1}\textbf{W}){\bf\Lambda}
    :={\bf\Phi}{\bf\Lambda},
    \label{eq:truncated-matrix}
\end{align}
where the columns of $\textbf{W}$ are eigenvectors and ${\bf\Lambda}$ is a diagonal matrix of the corresponding  eigenvalues $\lambda_{i}$, $i=1,\cdots,r$, and ${\bf\Phi}$ is called DMD modes \cite[Chap.~1]{kutz2016dynamic}.

Therefore, we have a reduced-order approximation:
\begin{align}
    \bfy_{k}\approx\sum_{j=1}^{r}{\bm\phi}_j\lambda_j^{k}b_j,
    \quad \Longrightarrow \quad 
    \bfY\approx{\bf\Phi}\bfD_{\bfb}{\bf\Xi}.
     \label{eq:DMD-approx}
    \end{align}
More precisely, we have the next compact form in detail:
\begin{align*}
   \bfY
 \!\approx\!
  \begin{smallmatrix}
   \underbrace{\begin{bmatrix}
      |    & \!\cdots\! & |\\
      {\bm\phi_1}& \!\cdots\! &   {\bm\phi_r}  \\
        |    & \!\cdots\! & |
    \end{bmatrix}}_{\Phi} 
    \underbrace{\begin{bmatrix}
        b_{1}  & &\\
        %& b_{2} & &\\
        & \ddots&\\
        & & b_{r}
    \end{bmatrix}}_{\bfD_{\bfb}}
    \underbrace{\begin{bmatrix}
        1  & \!\ldots\! & \lambda_{1}^{N-1}\\
       %  1 & \lambda_{2} & \ldots & \lambda_{2}^{N-1}\\
         \vdots  & \!\ddots\! &\vdots\\
          1 & \!\ldots\! & \lambda_{r}^{N-1}\\
    \end{bmatrix}}_{\bf\Xi}
    \end{smallmatrix},
\end{align*}
where ${\bf\Phi}\in{\mathbb{C}}^{p\times{r}}$ is the so-called \emph{spatial modes}, $\bfD_{\bfb}\triangleq{\mathrm{diag}}(b_{1},\ldots,b_{r})$ is the \emph{amplitudes} with diagonal form (resp., $\bfb\triangleq[b_{1},\ldots,b_{r}]$ denotes its vector case), and ${\bf\Xi}\in{\mathbb{C}}^{r\times{N}}$ indicates the Vandermonde matrix consisting of the eigenvalues to record \emph{temporal dynamics}.

%---------------------------------------------
%%%

\subsection{Sparsity-Promoting Dynamic Mode Decomposition}\label{subsec:method-II-SPDMD}
According to the above \eqref{eq:DMD-approx}, the DMD least-square error problem with $\ell_{0}$ ``norm''  regularization is written as % \cite{hastie2009elements}
\begin{align}
   \min_{\bfb\in{\mathbb{C}}^{r}}\|\bfY-{\bf\Phi}{\bfD_{\bfb}}{\bf\Xi}\|_{\text{F}}^{2} + \gamma\sum_{i=1}^{r}\iota({b_{i}\neq{0}}).
   %+ \lambda\|\bfb\|_{0},
  % \quad \text{s.t.}\quad \|\bfb\|_{0}\leq{s},
    \label{eq:least-square-prob-L0-regularization}
\end{align}
   % where $\|\bfb\|_{0}=\sum_{i}\iota({b_{i}\neq{0}})$
Here $\iota(\cdot)$ is an indicator function that measures the level of sparsity by counting the number of non-zero elements, that is, $1$ if $b_{i}\neq{0}$ and $0$ otherwise. A tunable parameter $\gamma\geq{0}$ denotes the  sparsity weight that makes the trade-off between the least-square error (i.e., accuracy) and the number of reduced amplitudes (i.e., model complexity).
%$s\in{\mathbb{R}}$ is tuning parameter that can be adjusted to achieve a desired level of sparsity.
As is well known, the program \eqref{eq:least-square-prob-L0-regularization} is a nonconvex optimization problem. However, it can be addressed using a convex surrogate, namely the $\ell_{1}$-regularized formulation \eqref{eq:regulaized-least-square-prob}, which is commonly referred to as \emph{sparsity-promoting} dynamic mode decomposition (SPDMD) \cite{jovanovic2014sparsity}.

Mathematically, we solve a \emph{regularized least squares error minimization problem} of \eqref{eq:least-square-prob-L0-regularization} as below
\begin{align}
   {\mathcal{J}}_{\gamma}(\bfb)\triangleq\min\limits_{\bfb}\|\bfY -{\bf\Phi}\bfD_{\bfb}{\bf\Xi}\|_{\text{F}}^{2} + 
   \gamma \sum_{i=1}^{r}|b_{i}|,
   %\gamma\|{\bfb}\|_{1},
    \label{eq:regulaized-least-square-prob}
\end{align}
where $\sum_{i=1}^{r}|b_{i}|$ indicates the sum of the absolute values of DMD amplitudes, or equivalently the $\ell_1$ norm $\|\bfb\|_{1}$.
Clearly, if $\gamma\to{0}$, the program~\eqref{eq:least-square-prob-L0-regularization} or \eqref{eq:regulaized-least-square-prob} reduces to the DMD computation \eqref{eq:least-square-prob}.
Problem \eqref{eq:regulaized-least-square-prob} is essentially a convex optimization, making it easily to obtain the optimal solution $\bfb^{\star}$ using the package, such as \texttt{CVX} in MATLAB \cite{grant2008cvx}.

In addition, the derivation $\|\bfY-{\bf\Phi}{\bfD_{\bfb}}{\bf\Xi}\|_{\mathrm{F}}^{2}$ of the sparse optimal amplitudes $\bfb^{\star}$, is obtained from the SPDMD algorithm, in fraction of normalized original data $\|\bfY\|_{\mathrm{F}}^{2}$,
then we take the performance loss $\Pi\%$ into account, evaluated by
\begin{align}
    \Pi\%:=100\sqrt{\frac{\mathcal{J}(\bfb^{\star})}{{\mathcal{J}}(0)}}=\frac{\|\bfY-{\bf\Phi}{\bfD_{\bfb}}{\bf\Xi}\|_{\mathrm{F}}}{\|\bfY\|_{\mathrm{F}}}\times{100}.
    \label{eq:loss-performance}
\end{align}

After solving the sparse optimization problem \eqref{eq:regulaized-least-square-prob} or \eqref{eq:least-square-prob-L0-regularization}, a sparse solution for the amplitudes $\bfb^{\star}$ is obtained, with cardinality $\mathrm{card}(\bfb^{\star}(\gamma)) = s$ (or $\ell_{0}$ norm $\|\bfb^{\star}(\gamma)\|_{0} = s$). This cardinality indicates the number of non-zero amplitudes corresponding to the most dominant modes, ordered by their magnitudes $|b_i|$ for $i = 1, \ldots, s$. Letting $m \triangleq \min\{s, r\}$, where $r$ is the SVD truncation rank in \eqref{eq:truncated-matrix}, we can consider a reduced-order model using only the selected $m$ modes.

\begin{figure*}[t!] 
    \centering
      \includegraphics[width=0.24\linewidth]{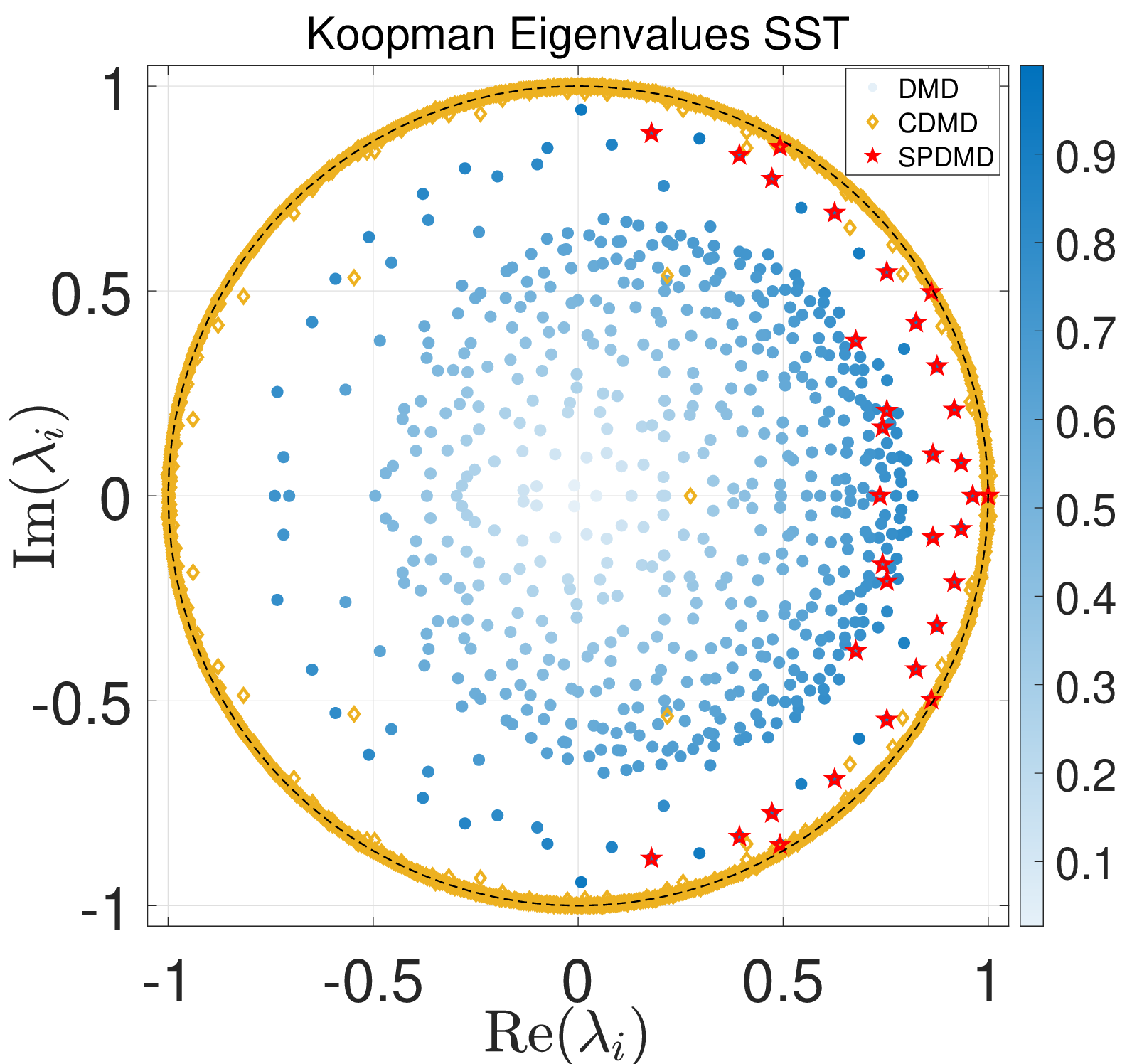}~
    \includegraphics[width=0.24\linewidth]{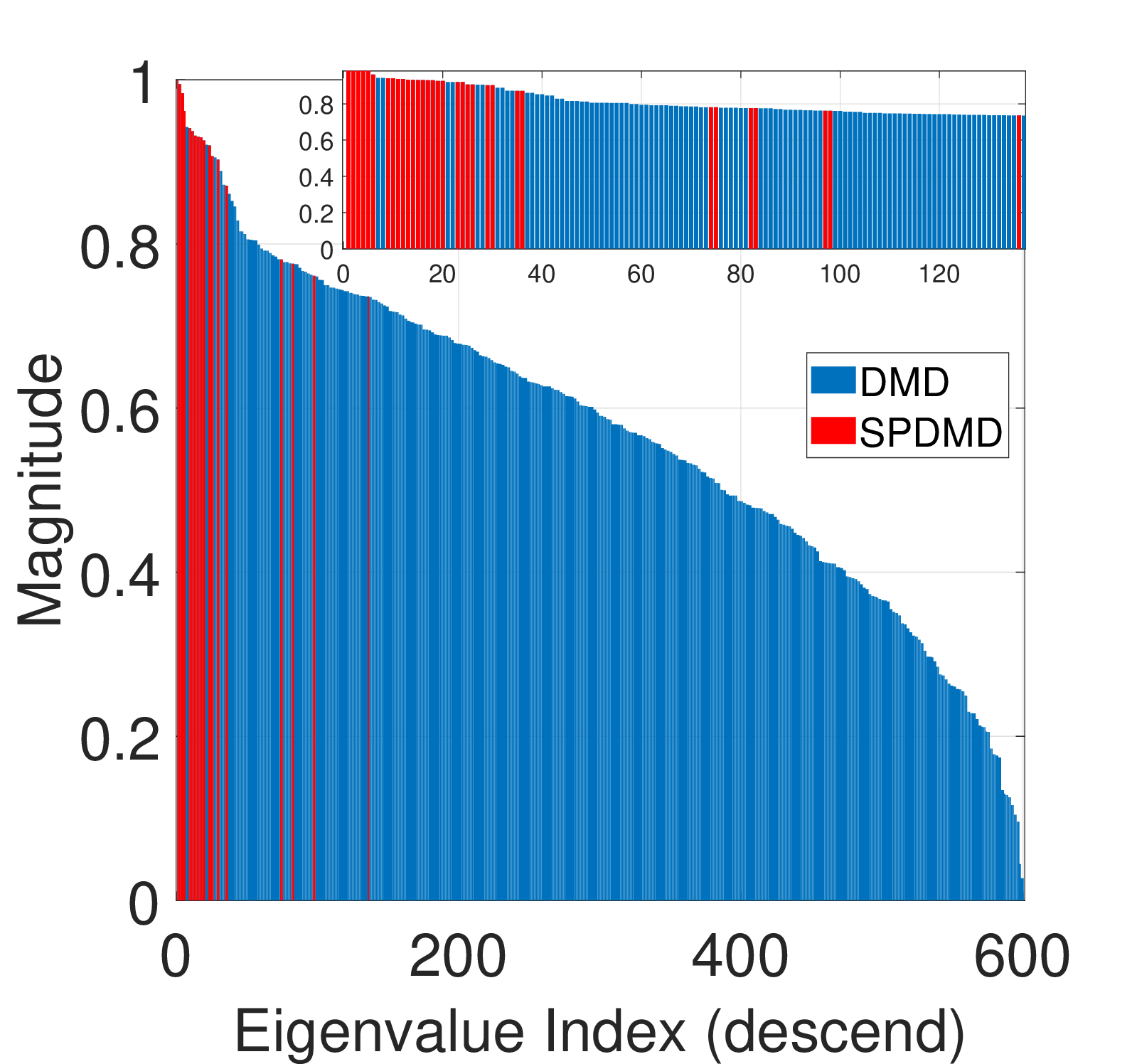}
      \includegraphics[width=0.24\linewidth]{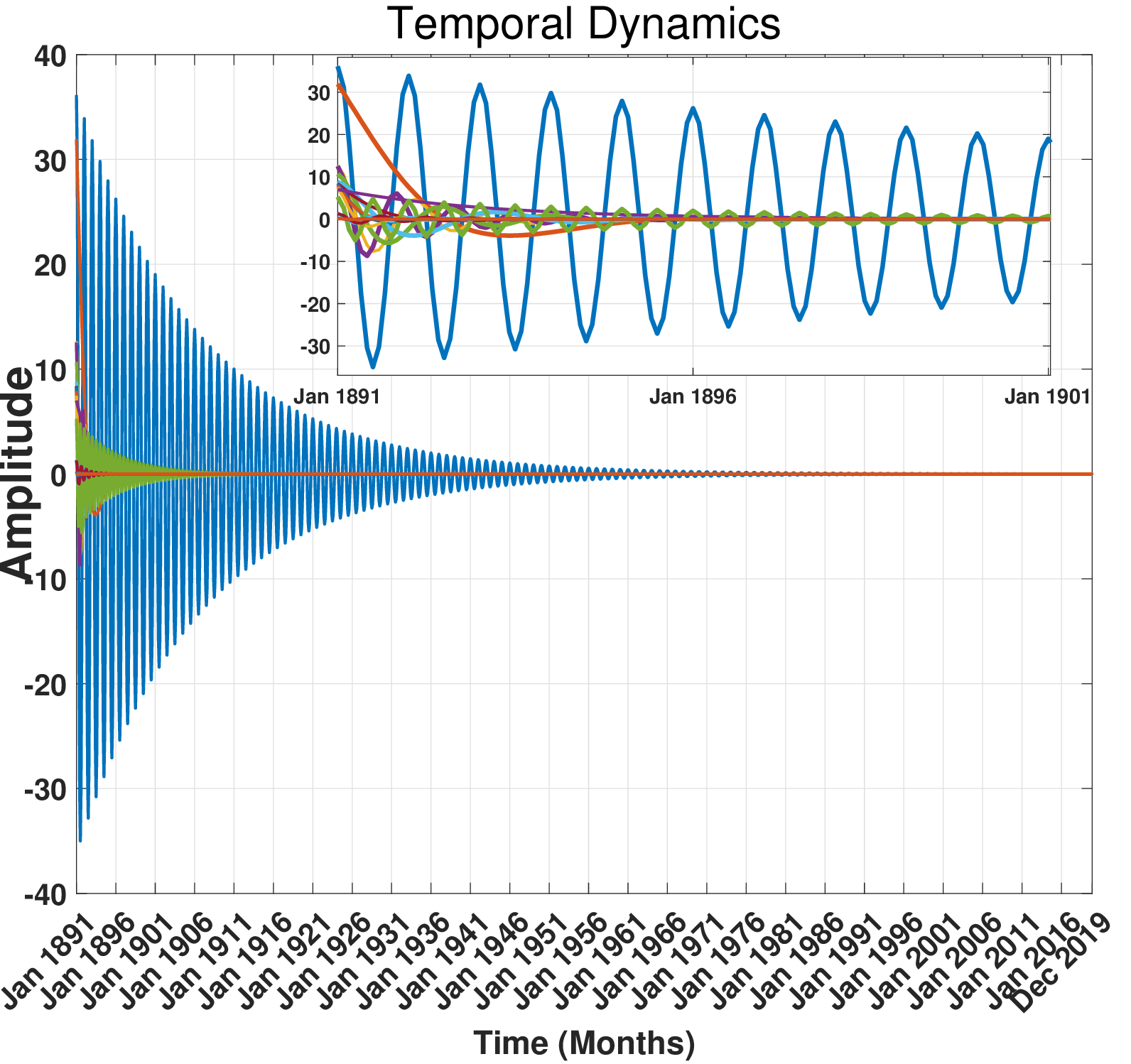}
      \includegraphics[width=0.25\linewidth]{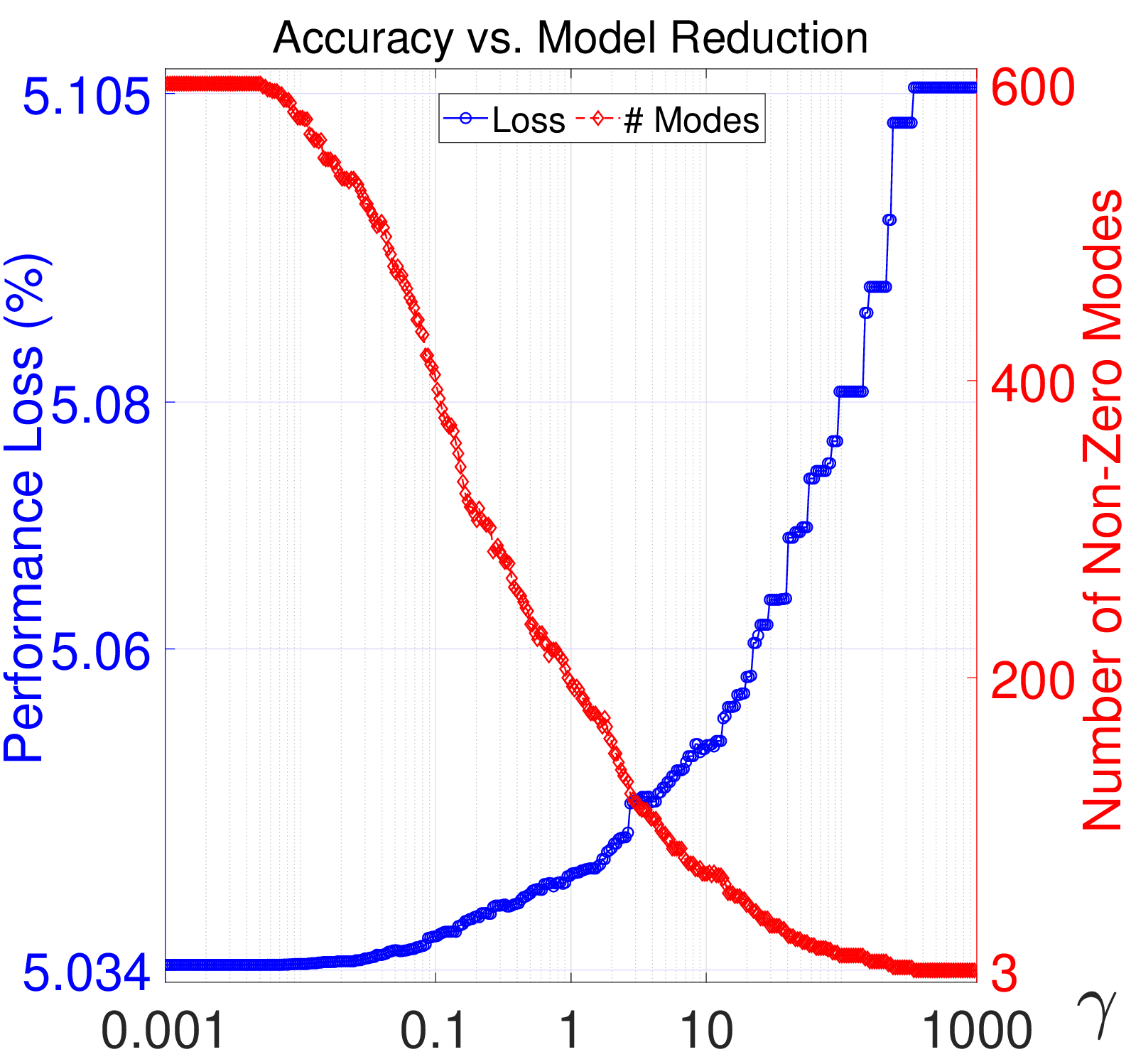}
    \caption{(left) The distribution of spectrum of Pacific monthly SST with all 600 eigenvalues obtained via DMD (blue), CDMD (yellow), and the $33$ dominant eigenvalues identified by SPDMD (red);
    (mid-I) the magnitude of the eigenvalues; 
    (mid-II) the time evolution of temporal dynamics by the selected modes;
    (right) the accuracy versus the model complexity.}
    \label{fig:Koopman-Eigvals-distributions}
 \end{figure*}

\begin{figure}[t!]
    \centering
    \includegraphics[width=\linewidth]{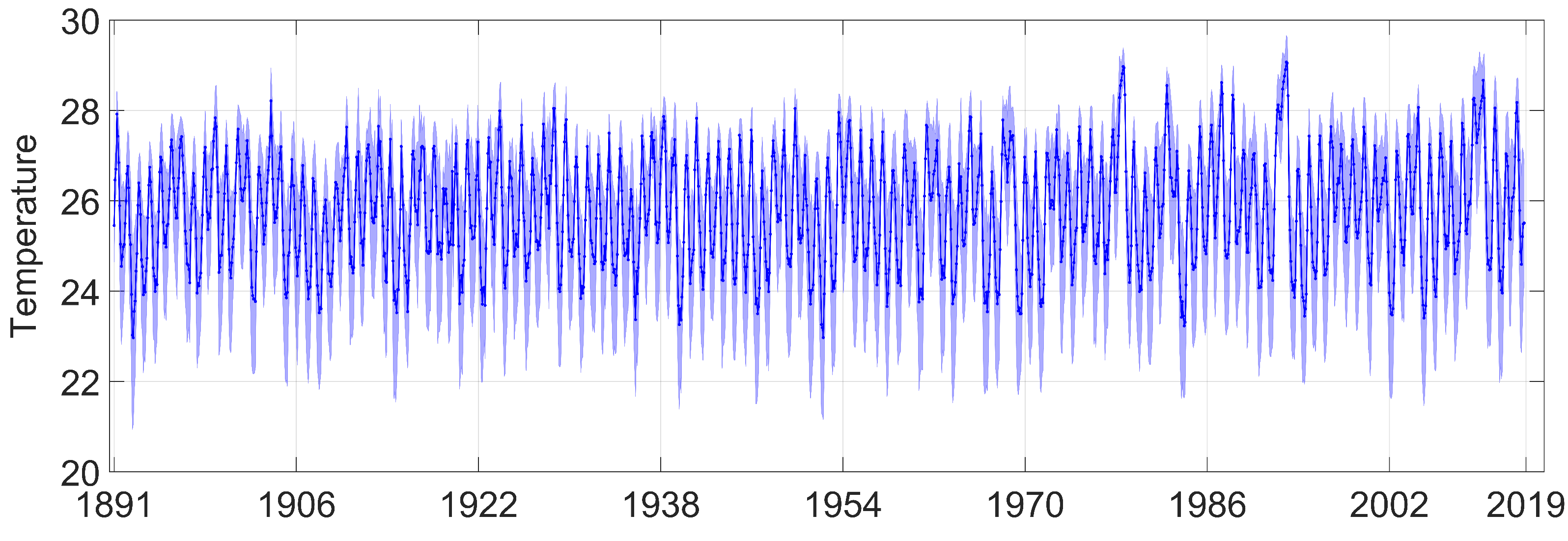}
    \caption{Ni\~{n}o-3 time-series monthly SST data in the Pacific: the shaded band shows one standard deviation, and the dot line denotes the mean-valued of the temperature.}
    \label{fig:SST-time-series}
    \end{figure}

\section{Applications: Sea Surface Temperature Field}\label{sec:SST_applications}
In this section, we apply KMD and SPDMD to real-world data-driven climate dynamics to extract the dominant Koopman modes that characterize Pacific SST variability and reveal underlying climate patterns.

\subsection{Data Source: Long-Term SST Data}
In this context, the (long-term) time-series dataset are obtained 
%from ERA5 and harvested 
from monthly SST over Ni\~{n}o-3 index\footnote{{\scriptsize \url{https://psl.noaa.gov/gcos_wgsp/Timeseries/Nino3}}}, taking only sea mask into account (without land data information), see in Fig.~\ref{fig:SST-time-series}; For more details, we refer the reader to the available data source and the related animation illustrating the time evolution of the SST data\footnote{SST data and animation are \href{https://drive.google.com/drive/folders/1_ie8o9FhhE9l8Zv_OJVIZ6IO6i3FBvBn?usp=drive_link}{available}.}.
%\cite{SSTAnimation}.
%\href{https://drive.google.com/drive/folders/1_ie8o9FhhE9l8Zv_OJVIZ6IO6i3FBvBn?usp=drive_link}{available.}

Besides, 
% the model is configured to cover the equatorial Pacific with a resolution of $0.25$\textdegree, and 
the chosen tropical region belt between the 5\textdegree{N} and 5\textdegree{S} in the direction of latitude and extends from 150\textdegree{W} to 90\textdegree{W} in the direction of longitude.
% The SST data are obtained from ERA5 and covered 
% We focus on the time-series dataset harvested from SST over Ni\~{n}o-3 index\footnote{\url{https://psl.noaa.gov/gcos_wgsp/Timeseries/Nino3}} and the chosen region extends from 5\textdegree{N} to 5\textdegree{S} (latitude) and from 150\textdegree{W} to 90\textdegree{W} (longitude).
More specifically, the dataset consists of monthly SST fields spanning from January $1891$ to December $2019$ for a total of $129$ years and corresponding long-term $1548$ months. %(i.e., $N=1547$). 
In other words, we collect the monthly SST data by stacking each snapshot into a  vector form as a column, then the rolling horizon data is 
%organized in a matrix:
%$\bfY=[\bfg_{1}(\bfx)~\bfg_{2}(\bfx) ~\cdots~\bfg_{N}(\bfx)]\in{\mathbb{R}}^{p\times{N}}$
 \begin{align}
   %   \bfY:=
      % \begin{bmatrix}
      %   \bfg_{1}(\bfx)& \bfg_{2}(\bfx) & \cdots & \bfg_{N}(\bfx)
      % \end{bmatrix}\in{\mathbb{R}}^{p\times{N}},
  %  \begin{bmatrix}
       %  \bfy_{0} & \bfy_{2} & \cdots & \bfy_{N-1}
   %  \end{bmatrix} \in{\mathbb{R}}^{p\times{N}},
   \{\bfy_{0},\bfy_1,\ldots,\bfy_{N}\},
  \label{eq:SST-snapshot-matrix}
\end{align}
where the vector $\bfy_{k}\in{{\mathbb{R}}}^{600\times{1}}$ of spatial length $p=600$ denotes the system at different time steps $k=0,1,\ldots,1547$ (that is, $N=1548$). Let $\bfY^{+}$ be the one time delay-embedding (or time-shifted) copies of the zero-lag time data matrix of $\bfY$ in \eqref{eq:data-snapshots}.
Each data point contains monthly SST data measured on a snapshot size of $10\times{60}=600$ grids, with $60$ grid points along the longitude and $10$ grid points along the latitude. %(see the available {\href{https://drive.google.com/file/d/1Rx9mtVROqT995Y5EgYwlsbXGa4qJB-uC/view?usp=drive_link}{animation}}). 
Therefore, in general each grid point implies a degree of the freedom, indicating a high dimensional climate system problem.
%Namely, the original SST data is $\bfY\in{\mathbb{R}}^{600\times{1548}}$. 

  \begin{table}[t!] 
\centering 
%\scriptsize  % Reduce font size
\caption{The first $11$ key modes ordered by amplitudes.}
\label{tabel:mode-norm-e-folding-periodic} 
\begin{tabular}{@{\extracolsep{3pt}} c c c c c} 
\hline 
Mode & Amps & Norm & $e$-folding time & Period (mths)\\ 
$i$  & $|b_{i}|$ & $|\lambda_{i}|$ & $-{1}/{|\mathrm{Re}(\log\lambda_{i})|}$ & ${2\pi}/{\mathrm{Im}(\log\lambda_{i})}$\\
\hline 
\rowcolor{lightgray}
1 (45) & 641.79 & 1.00 & Inf & Inf \\
3 (41) & 36.14  & 0.99 & 186.97 & 12.00 \\
5 (66) & 31.94  & 0.94 & 15.48 & 73.15 \\
7 (60) & 12.64  & 0.93 & 13.03 & 13.26 \\
9 (52) & 12.52  & 0.93 & 13.76 & 10.01 \\
11 (62) & 10.69  & 0.93 & 13.92 & 18.12 \\
13 (63) & 8.80  & 0.94 & 16.47 & 27.84 \\
15 (72) & 8.39 & 0.87 & 7.24 & 53.98 \\
17 (248) & 8.33 & 0.78 & 4.05 & 23.29 \\
19 (294) & 7.91 & 0.76 & 3.66 & 28.37 \\
21 (145) & 7.46 & 0.78 & 3.94 & 12.33 \\
\hline
\end{tabular} 
\end{table}

\subsection{Experiment~1: Monthly SST Data Field}\label{subsec:experiment-MSST}
We first explore climate dynamics, focusing on high-resolution (i.e., one data point per month) intraseasonal SST variability by studying Koopman modes and their month-to-month changes. This variability plays a key role in simulating tropical phenomena, such as MJO, which is characterized by a phase evolution over a $30–90$ day period \cite{lintner2023identification}.
The spectral distribution of the linear operator on monthly SST is shown in Fig.~\ref{fig:Koopman-Eigvals-distributions} (left). In this figure, the ``blue dots'' represent eigenvalues estimated using the DMD algorithm described in 
%\eqref{eq:least-squre-solution} and 
\eqref{eq:truncated-matrix}, the ``yellow diamonds'' correspond to the companion-based DMD (CDMD) method \cite{rowley2009spectral}, and the ``red stars'' denote eigenvalues obtained through SPDMD \cite{jovanovic2014sparsity} as formulated in \eqref{eq:regulaized-least-square-prob}. % by the order of the absolute values of the amplitudes. 
It is worth noting that for such a flat data matrix, the CDMD is particularly useful for studying oscillatory behavior (e.g., MJO event), as the spectrum of the companion matrix is predominantly distributed along the unit circle. In contrast, both standard DMD and SPDMD tend to place the eigenvalues strictly inside the unit circle.
%, as see in Fig.~\ref{fig:Koopman-Eigvals-distributions}. 
Nonetheless, all three methods (DMD, CDMD, and SPDMD) consistently indicate that the modes are stable, as verified by Koopman eigenvalues with magnitude or absolute value less than one in Fig.~\ref{fig:Koopman-Eigvals-distributions} (mid-I), implying that the monthly SST data-embedded climate system enjoys a \emph{stationary state}.
Additionally, Table~\ref{tabel:mode-norm-e-folding-periodic}
presents the first $11$ dominant Koopman modes, ordered by the magnitude of their amplitudes (along with their corresponding original location numbers). 

\begin{figure*}[t!] 
    \centering
          \includegraphics[width=0.24\linewidth]{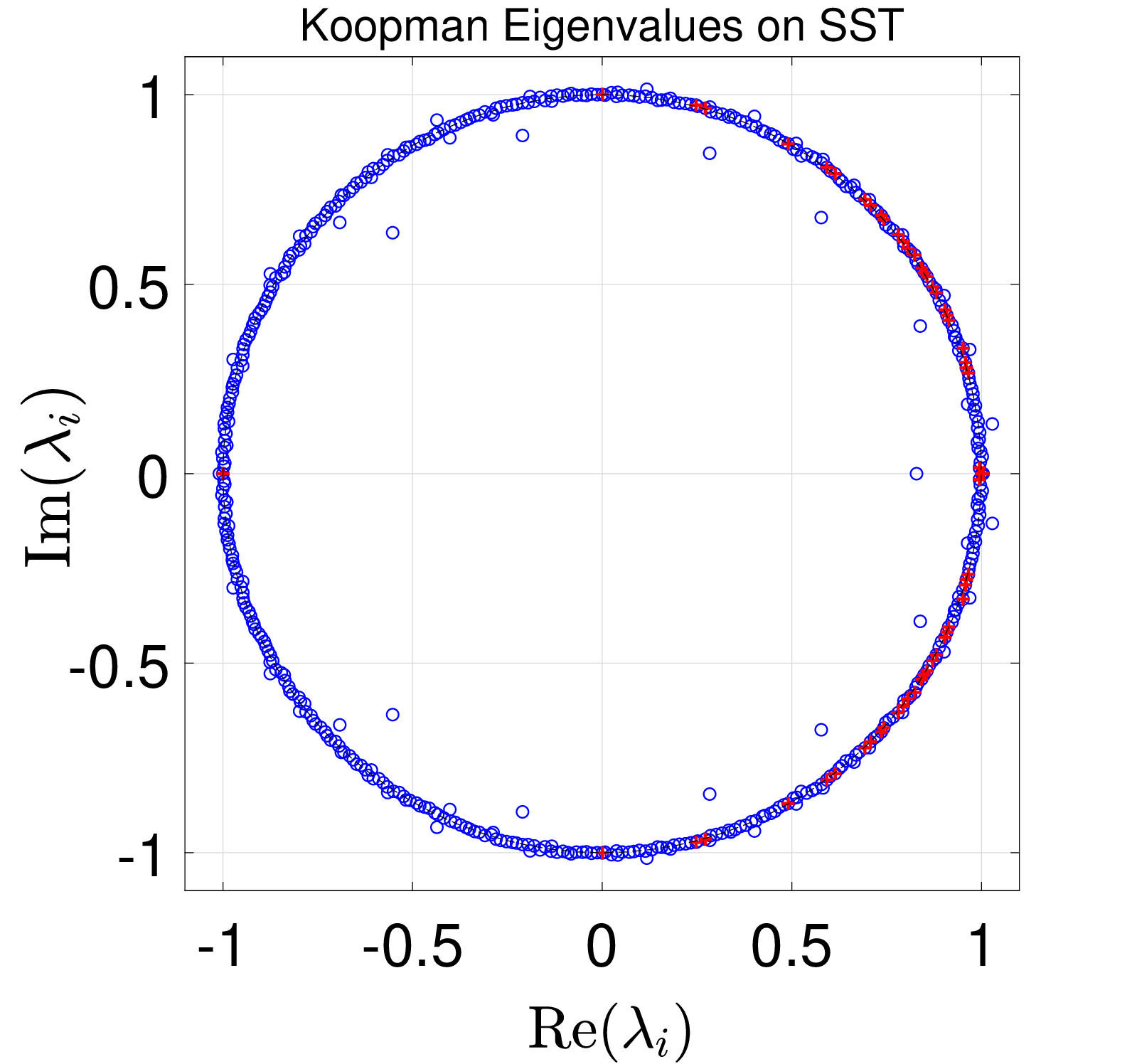}
    \includegraphics[width=0.24\linewidth]{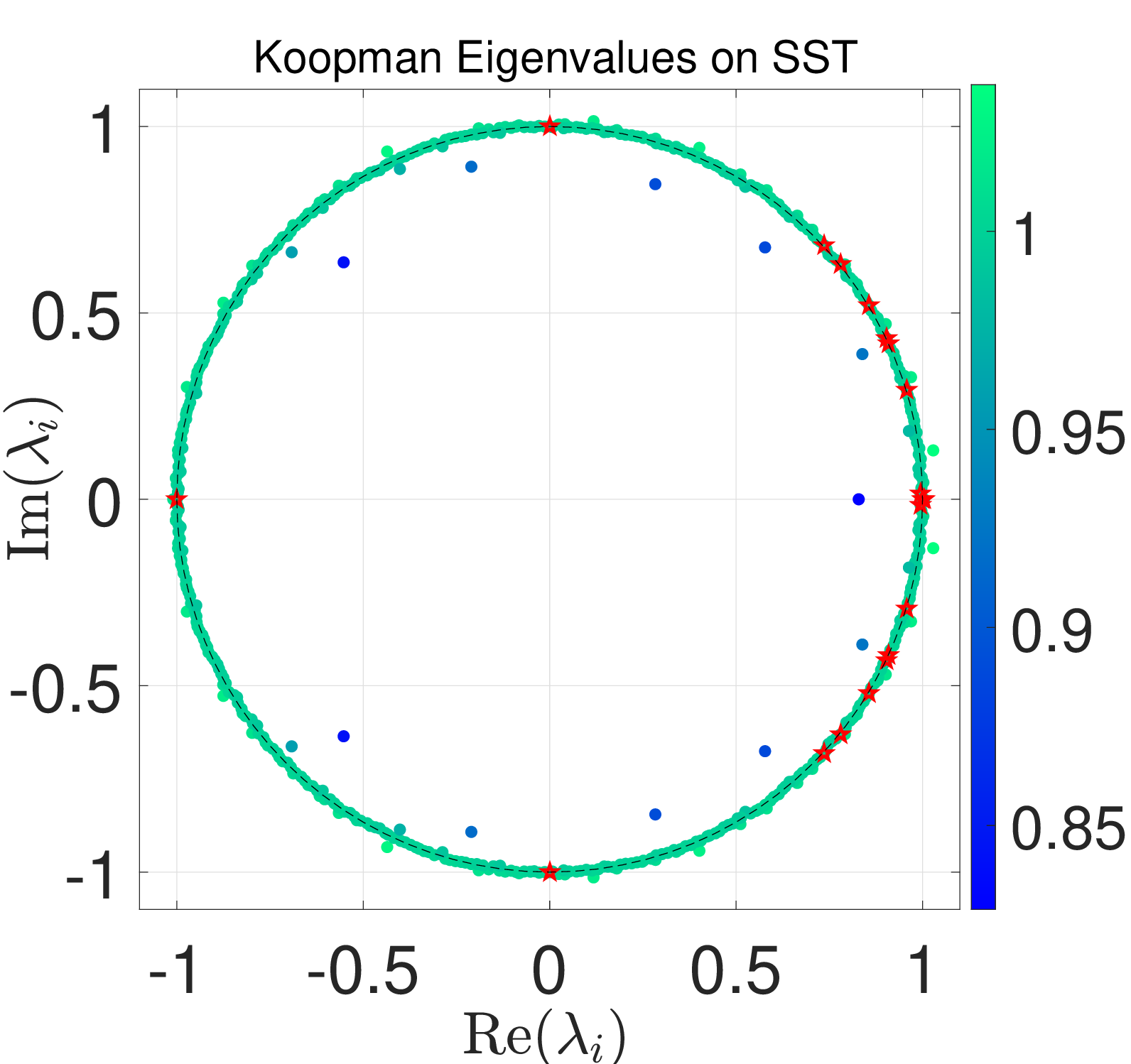}
      \includegraphics[width=0.245\linewidth]{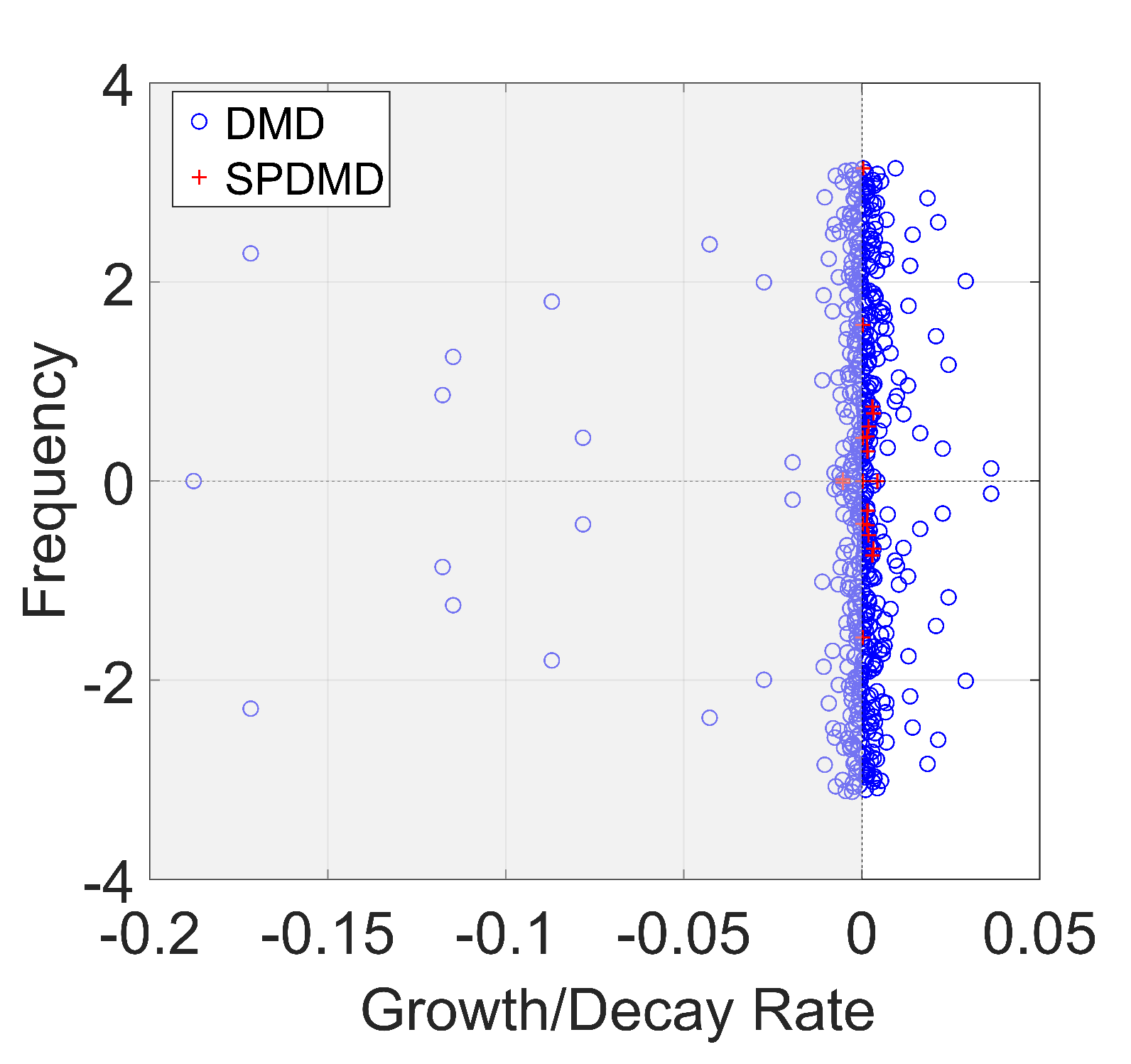}
        \includegraphics[width=0.24\linewidth]{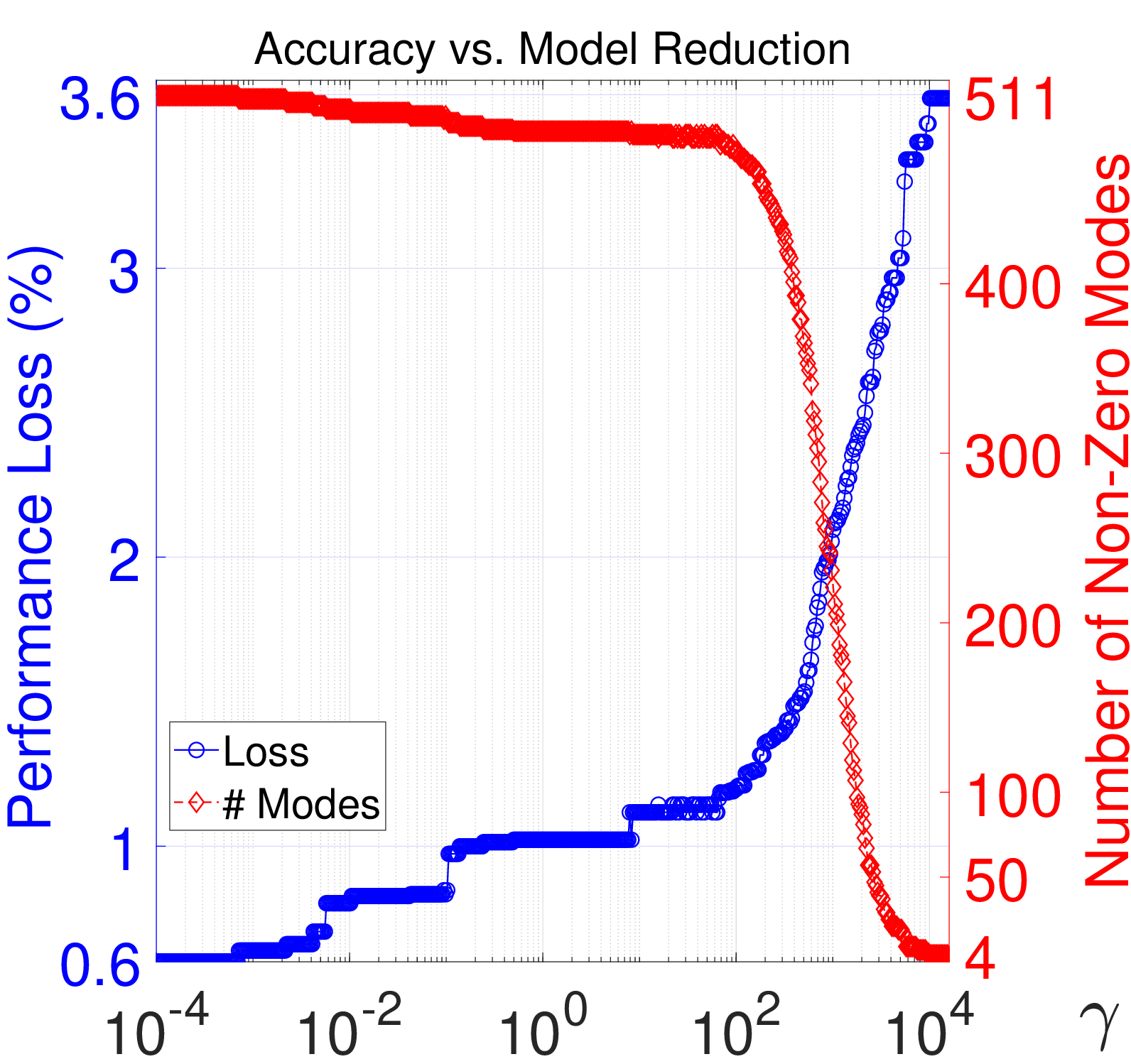} 
    \caption{(left) Koopman eigenvalues for the seasonal cycle of the SST data field, including all $515$ modes and the selected $57$ eigenvalues; (mid-I) The $19$ eigenvalues selected by SPDMD; (mid-II) the growth or decay rate versus the frequency; (right) the balance between accuracy and model reduction under different sparsity weights $\gamma\in[0.0001,16000]$.}
    \label{fig:Koopman-eigvals-quarterly-SST-case}
 \end{figure*}

By promoting sparsity on the amplitudes, it consists entire spectrum of $r=600$ modes, reducing it to a smaller number of $m=33$ modes. This reduction is significant as it enables us to capture the leading modes of the high-dimensional climate system, allowing for a more focused analysis through a low-dimensional system representation. Fig.~\ref{fig:Koopman-Eigvals-distributions} (mid-II) illustrates the time evolution of the real part of the temporal dynamics governed by ${\mathrm{Re}}(\lambda_{j}^{t}b_{j})$, corresponding to the first $17$ leading modes (conjugate pairs) ordered by the magnitude of their amplitudes $|b_j|$. In the case of complex conjugate pairs, only one spatiotemporal mode is shown. These dynamics are governed by the Koopman eigenfunctions, whose temporal evolution eventually decays to steady-state in zero. %as their eigenvalues are less than one in magnitude.

\begin{table}[b!] 
\centering 
%\scriptsize  % Reduce font size to help fit in column
%\footnotesize
\small
  \caption{Model reduction with the number of modes, the cost, and the performance loss in seasonal cycle of SST.} 
  \label{tabel:SST-quarter} 
\begin{tabular}{@{\extracolsep{5pt}} c c c c} 
\\[-1.8ex]\hline 
\hline \\[-1.8ex]
Weight $\gamma$  & Modes $\#i$ & Cost $\mathcal{J}(\mathbf{b}^{\star})$ & Loss $\Pi\%$ \\ 
\hline \\[-1.8ex] 
 0.0001 &  511 &  2.2308$\times10^{4}$ & 0.6010\\
 0.0105 &  501 &  1.1602$\times10^{5}$ & 0.8277\\
 0.5503 &  490 &  1.8414$\times10^{5}$ & 1.0224\\
 8.5573 & 488 &  2.2331$\times10^{5}$ & 1.1172\\
  \rowcolor{lightgray}
100.24 &  478 &  2.3250$\times10^{5}$ & 1.1983\\
691.05 &  303 &  3.4864$\times10^{5}$ & 1.8255\\
1714.3 &  107 &  4.7381$\times10^{5}$ & 2.3815\\
4948.2 &  19 &   7.7276$\times10^{5}$ & 3.0350\\
5757.2 & 9 &  8.6025$\times10^{5}$ & 3.3761\\
16000 &  4 &  1.6447$\times10^{6}$ & 3.5880\\[0.5ex] 
\hline
\end{tabular} 
\normalsize 
\end{table}

Fig.~\ref{fig:Koopman-Eigvals-distributions} (right) illustrates the trade-off between accuracy (evaluated by performance loss) and model complexity (measured by the number of non-zero modes) as analyzed via SPDMD in the monthly SST. 
By tuning the sparsity weight $\gamma\in[0.001,1000]$ with $350$ grids, we observe that the deviation in least squares error is not negligible, with a loss of $\Pi\%=5.034\%$. 
Also, the number of modes gradually decreases from $600$ to $3$ as the weight increases.

\begin{table}[b!] 
\centering 
\caption{The first $10$ key modes ordered by amplitudes.} 
\label{tabel:mode-norm-eigvals-periodic-quarter} 
\resizebox{0.98\columnwidth}{!}{%
\begin{tabular}{@{\extracolsep{2pt}} c c c c c} 
\hline 
Mode & Amps & Norm & Eigenvalue & Period (season)\\ 
$i$  & $|b_{i}|$ & $|\lambda_{i}|$ & $\lambda_{i}$ & ${2\pi}/{\mathrm{Im}(\log\lambda_{i})}$\\
\hline 
\rowcolor{lightgray}
1 (412) & 1058.52 & 1.00 & 1.0003 + 0.0000i & Inf \\
3 (445) & 24.47   & 1.00 & 0.0005 -- 1.0003i & 4.00 \\
5 (505) & 3.19    & 1.00 & -1.0004 + 0.0000i & 2.00 \\
\rowcolor{lightgray}
7 (417) & 1.73    & 0.99 & 0.9946 + 0.0153i  & 407.19 \\
9 (180) & 1.49    & 1.00 & 0.8563 -- 0.5203i & 11.51 \\
11 (242)& 1.23    & 1.00 & 0.9035 -- 0.4320i & 14.09 \\
13 (7)  & 0.36    & 1.00 & 0.7361 + 0.6813i  & 8.41 \\
15 (335)& 0.36    & 1.00 & 0.9576 + 0.2934i  & 21.31 \\
17 (20) & 0.29    & 1.00 & 0.7803 -- 0.6307i & 9.24 \\
19 (264)& 0.21    & 1.00 & 0.9088 -- 0.4180i & 14.57 \\
\hline
\end{tabular}
}
\end{table}

%%=============Spatial modes for quaterly 
\begin{figure*}[t!]
   \centering
    \includegraphics[width=0.32\linewidth]{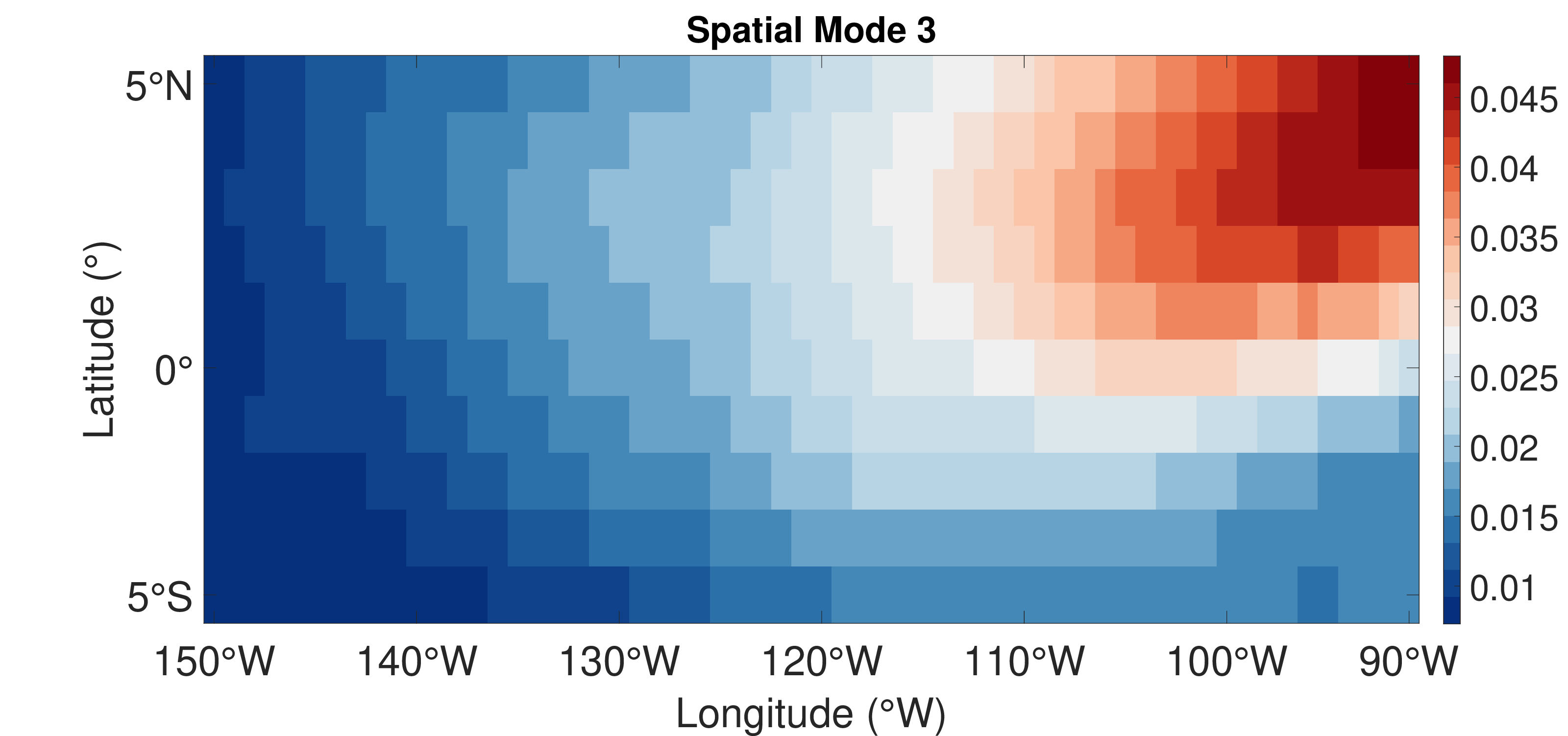}
     \includegraphics[width=0.32\linewidth]{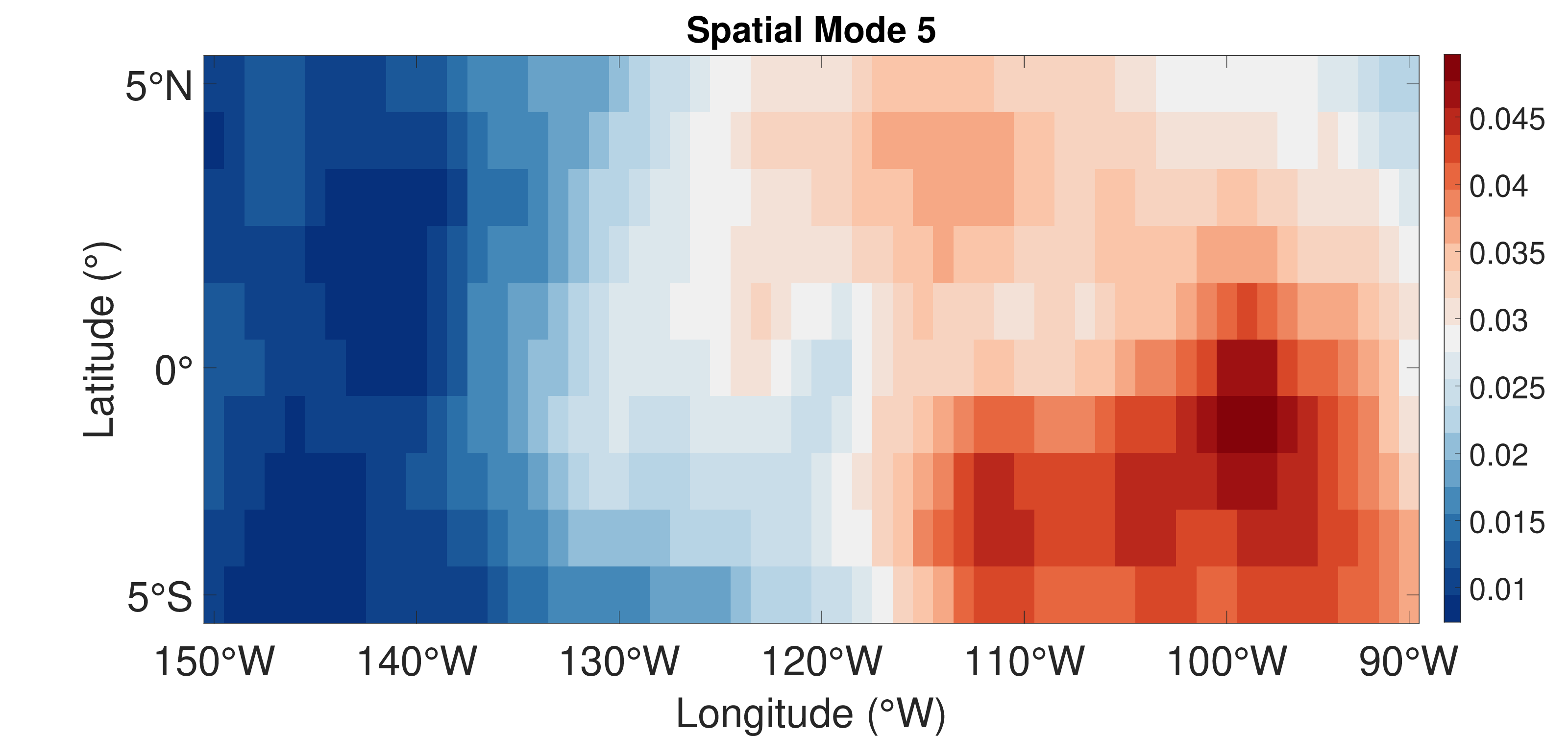}
    \includegraphics[width=0.32\linewidth]{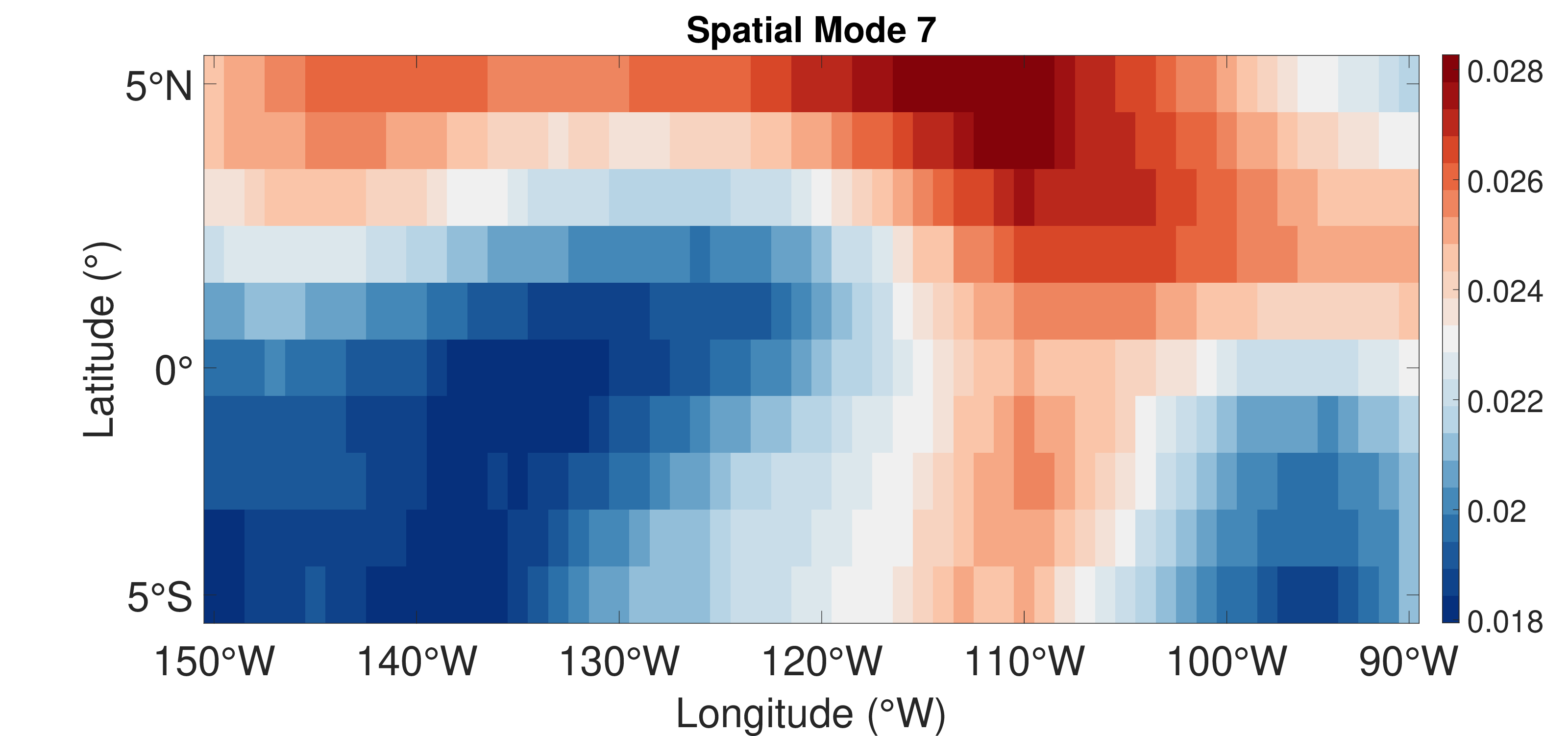}
       \includegraphics[width=0.32\linewidth]{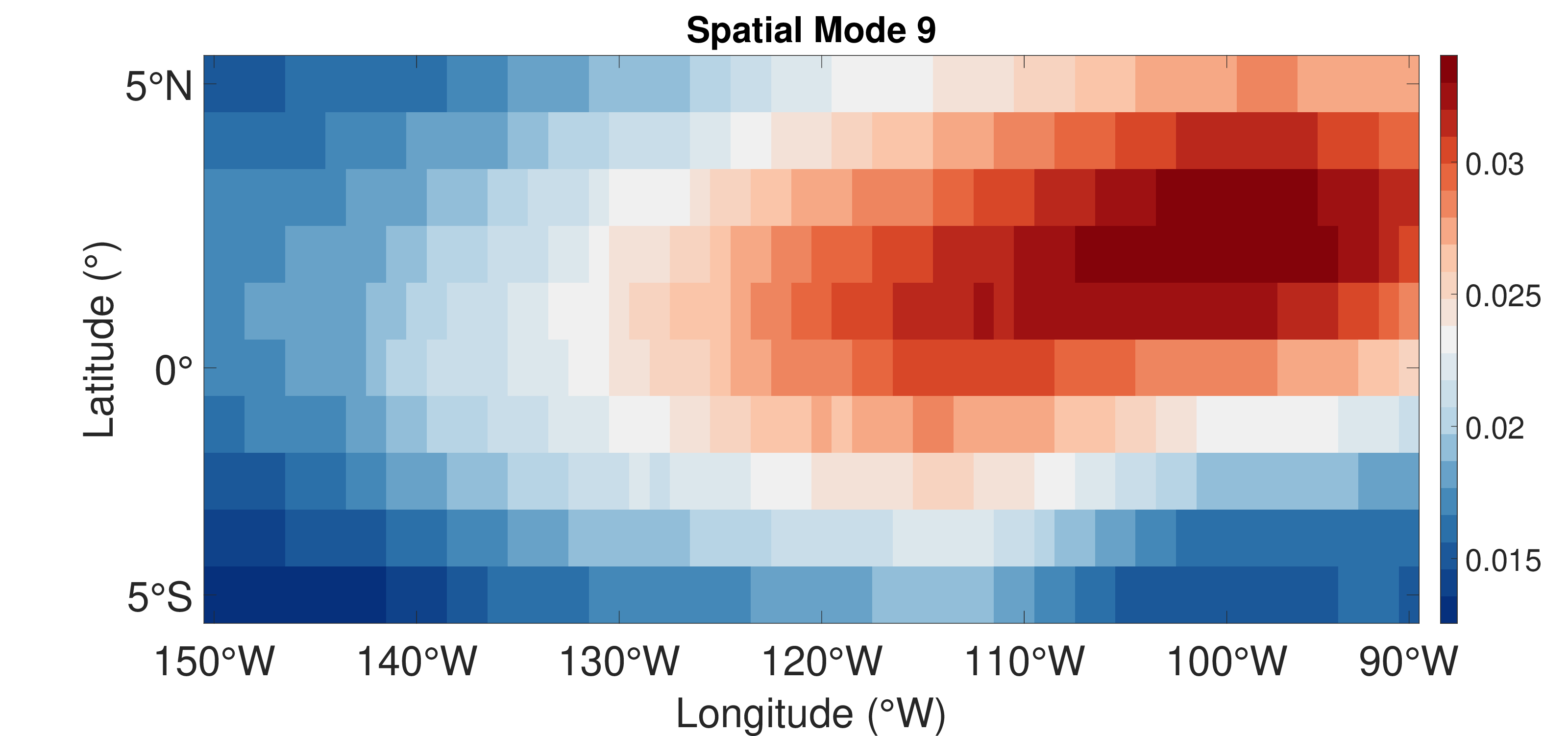}
        \includegraphics[width=0.32\linewidth]{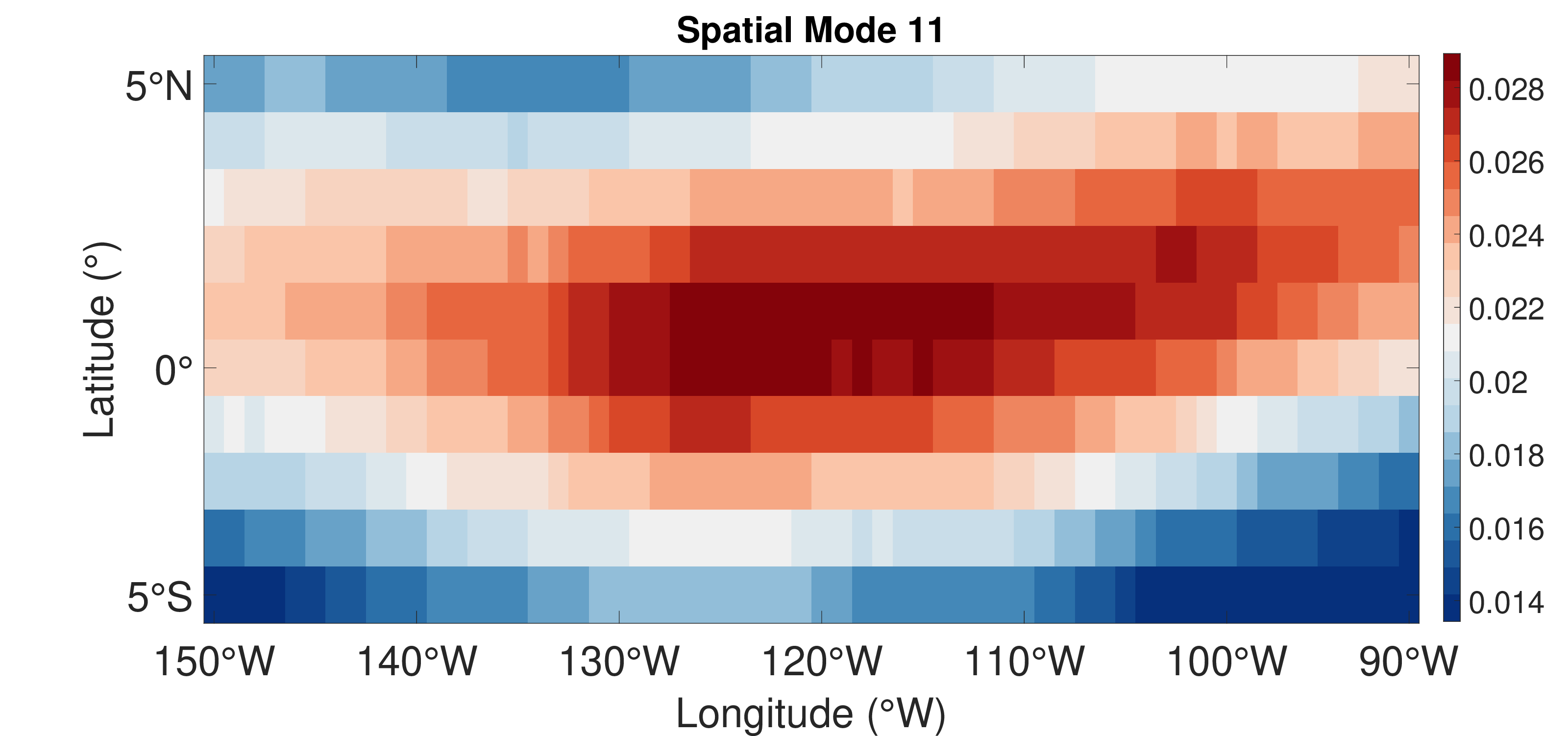}
       \includegraphics[width=0.32\linewidth]{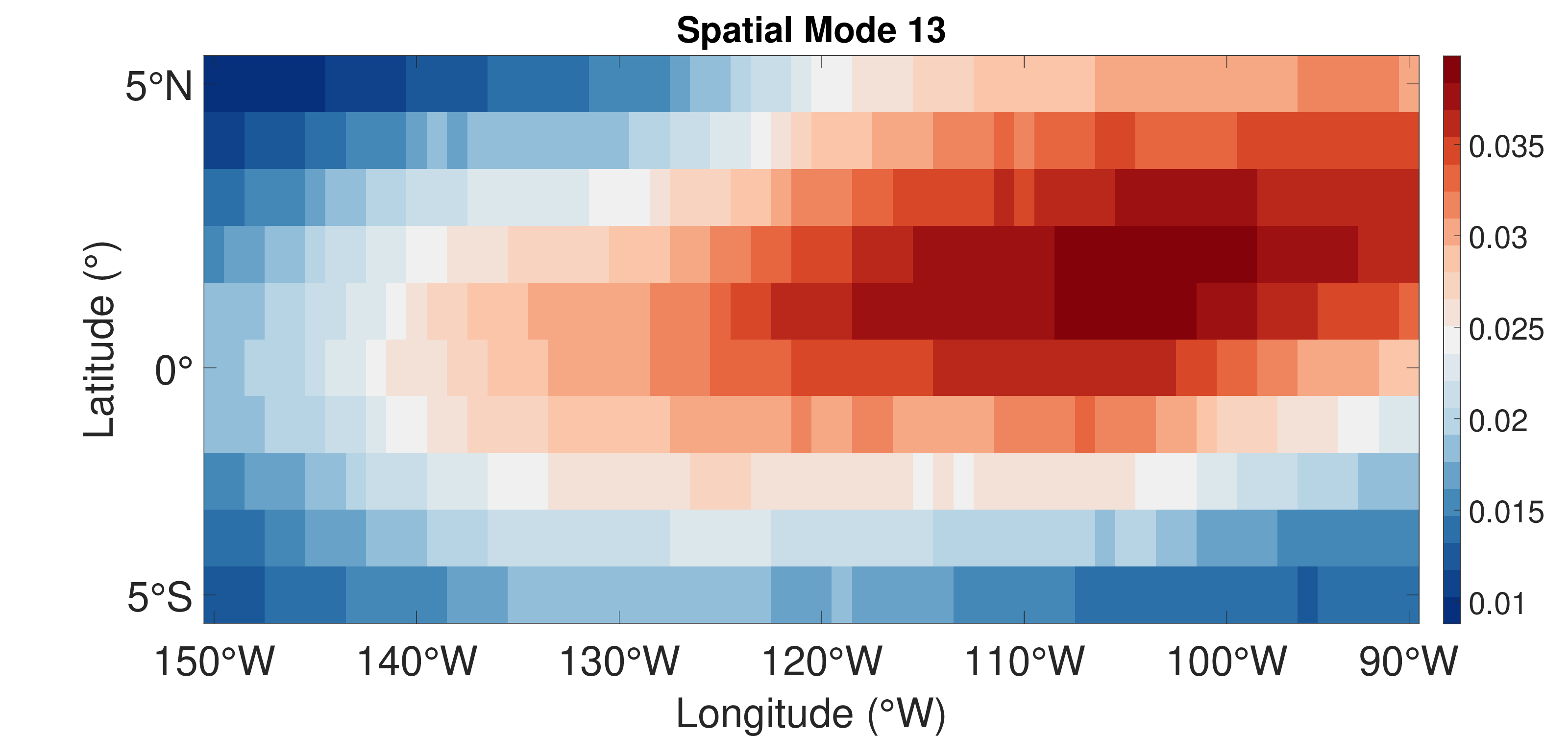}
          \includegraphics[width=0.32\linewidth]{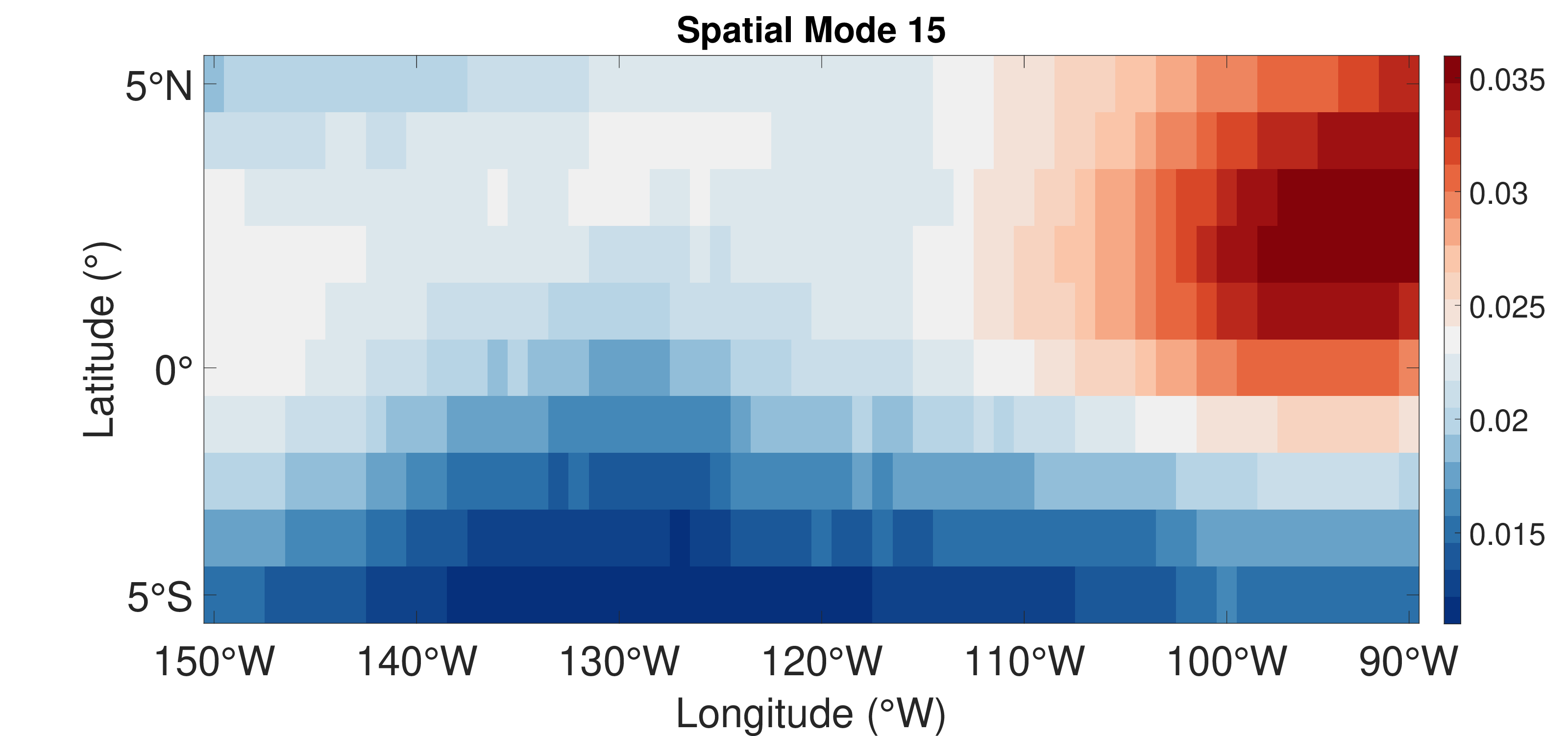}
        \includegraphics[width=0.32\linewidth]{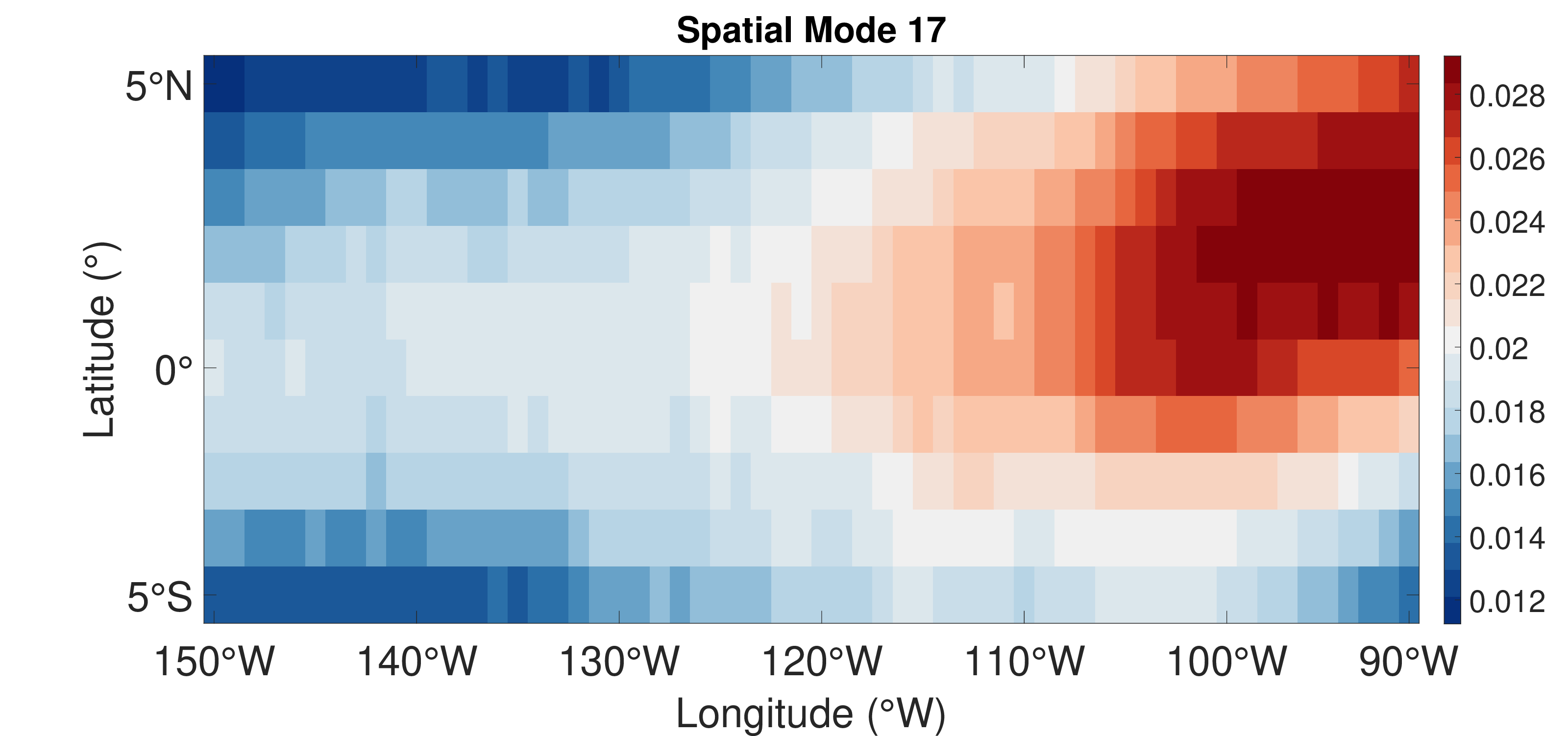}
        \includegraphics[width=0.32\linewidth]{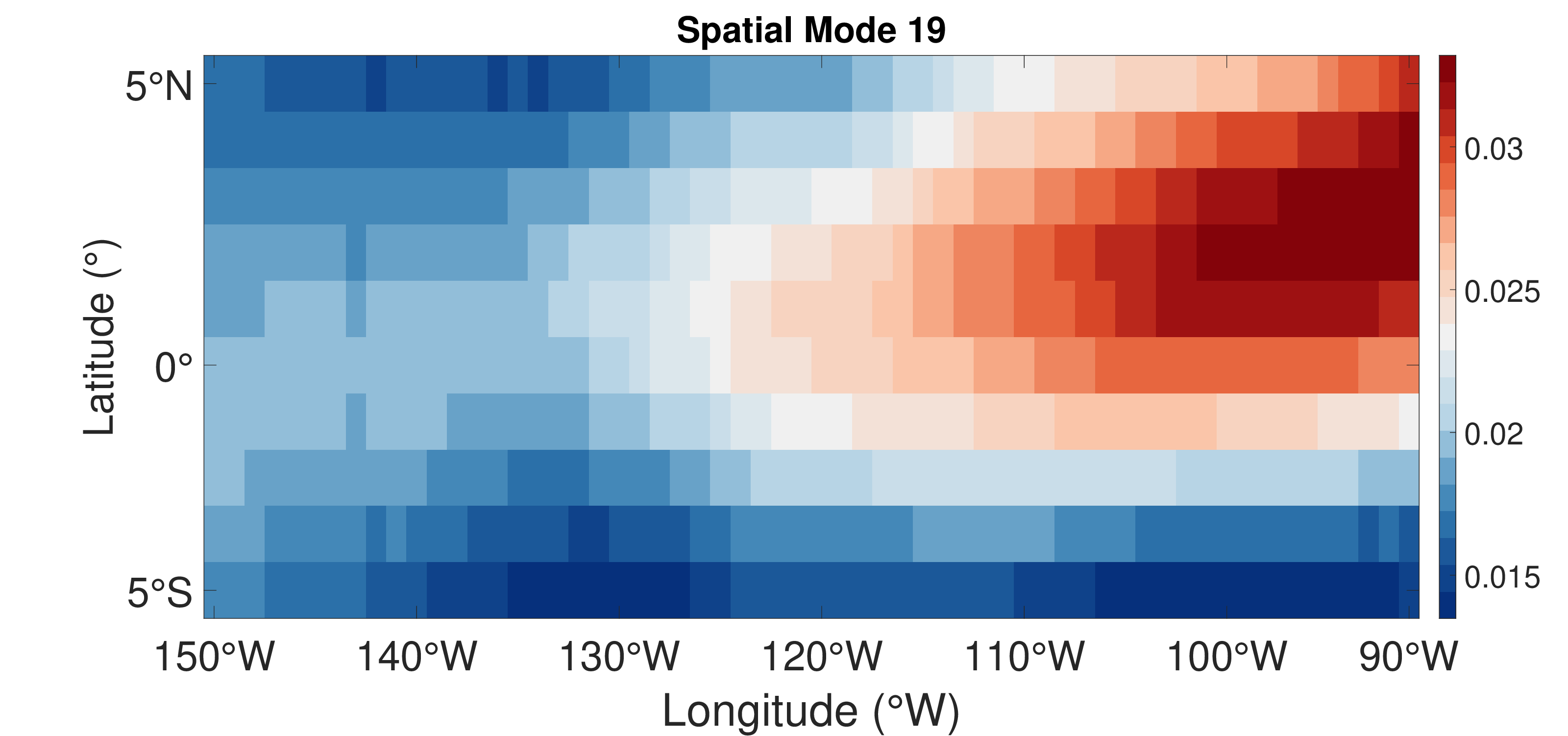}
   \caption{Spatial Modes for seasonal cycle of SST data.}
   \label{fig:spatial-modes-quarter}
 \end{figure*}
 
\subsection{Experiment~2: Seasonal Cycle of SST Data Field}\label{subsec:experiment-QSST}
We now discuss the seasonal cycle of SST data field by first organizing the $3$-month data matrices into a vector field. Thus,
the snapshot spans the spatial points, encompassing a medium resolution time scale (one data point per season, typically averaging $3$ months of grid data), and forms a high-dimensional vector $\bfy_k\in{\mathbb{R}}^{1800\times{1}}$ 
with $p = 1800$. In this case,
the data for each quarter is then aggregated by stacking the spatial points from each season into a single vector.
Consequently, the number of snapshots becomes to $N=515$, resulting in a \emph{tall-and-skinny} data matrix $\bfY\in{\mathbb{R}}^{1800\times{515}}$.
This data matrix $\bfY$ is transformed into four seasonal cycles, helping to eliminate seasonal variations, such as changes in temperature and precipitation factors.
In general, quarterly SST data is good for capturing the ENSO 
\cite{froyland2021spectral}, as it concerns on a seasonal-to-interannual timescale, with its impacts often becoming most apparent over several months to a year.

\begin{figure}[!h]
    \centering     
    \includegraphics[width=\linewidth]{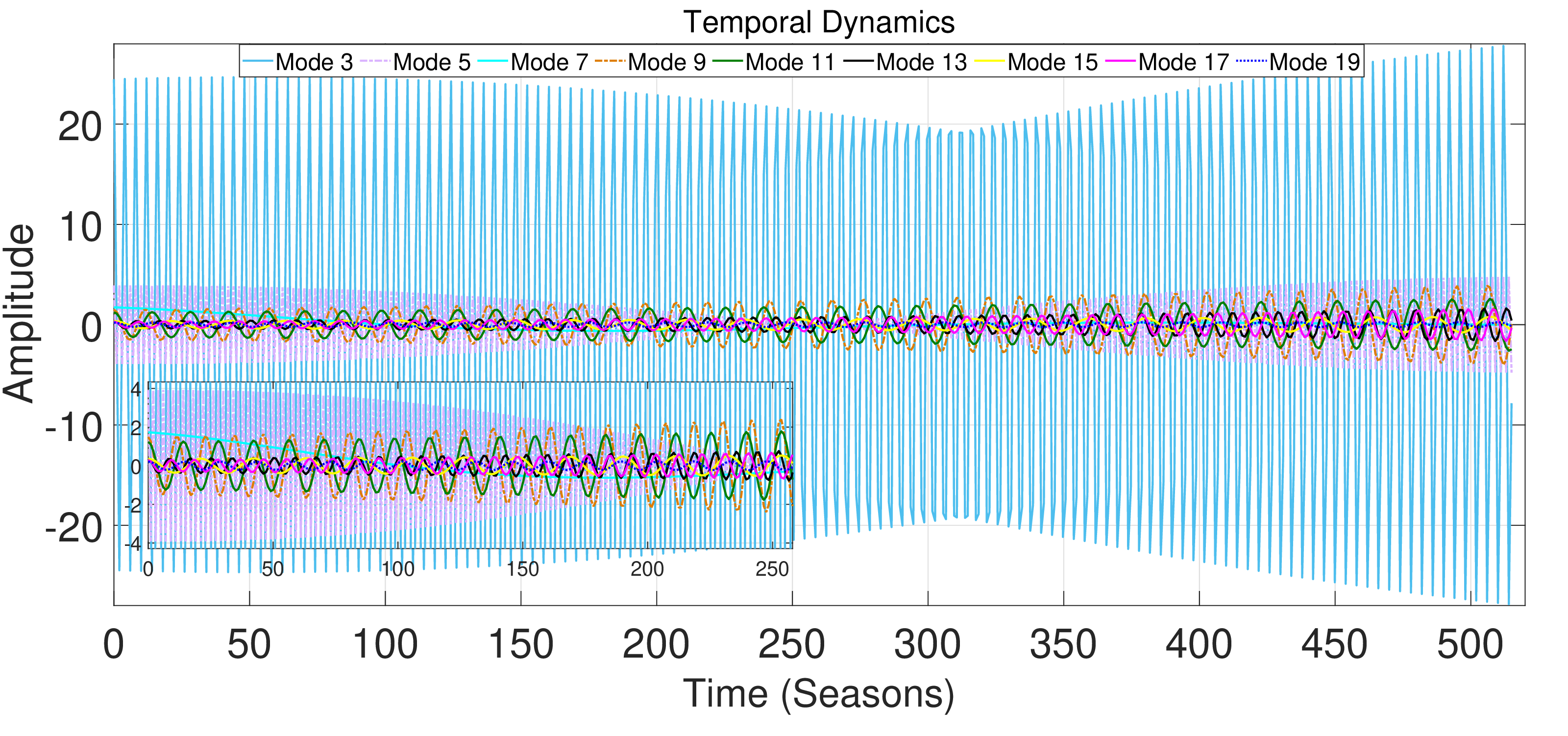}
    \caption{The real part of selected $9$ temporal modes show the oscillations with periodic behavior in seasonal SST.}
    \label{fig:quarter-SST-temporal-dynamics}
 \end{figure}

\begin{figure*}[t!]
    \centering
\begin{subfigure}{0.32\textwidth}
       \centering
      \includegraphics[width=\linewidth]{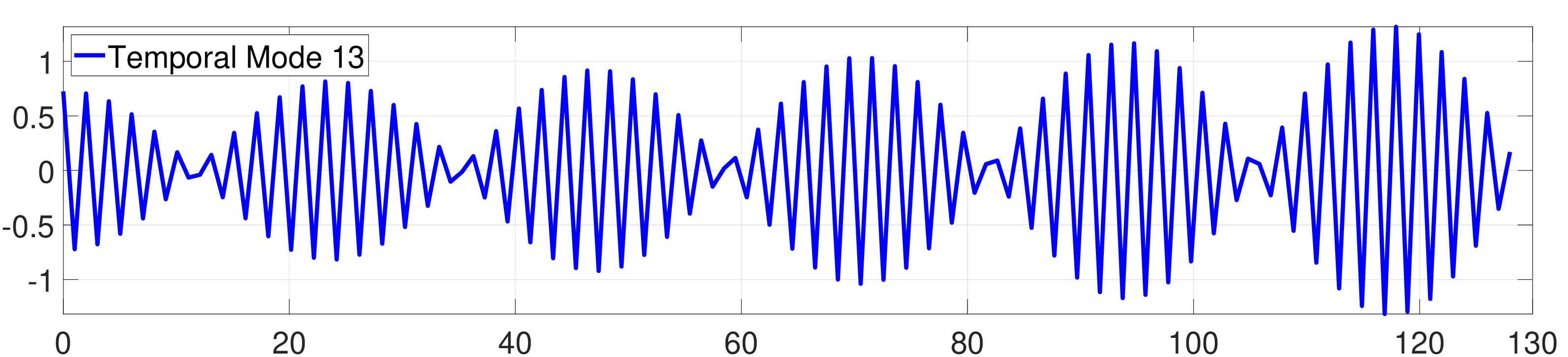} \\
      \includegraphics[width=\linewidth]{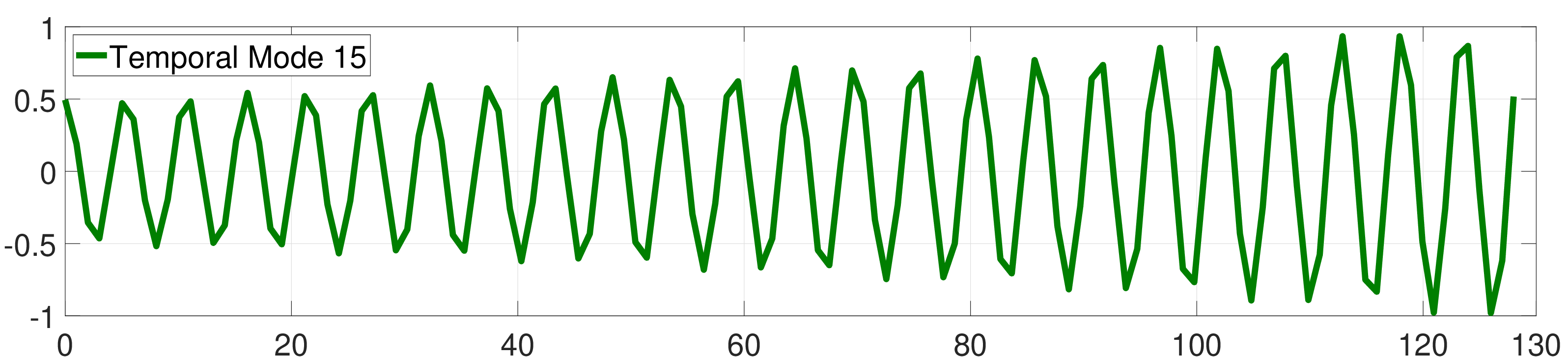}\\
      \includegraphics[width=\linewidth]{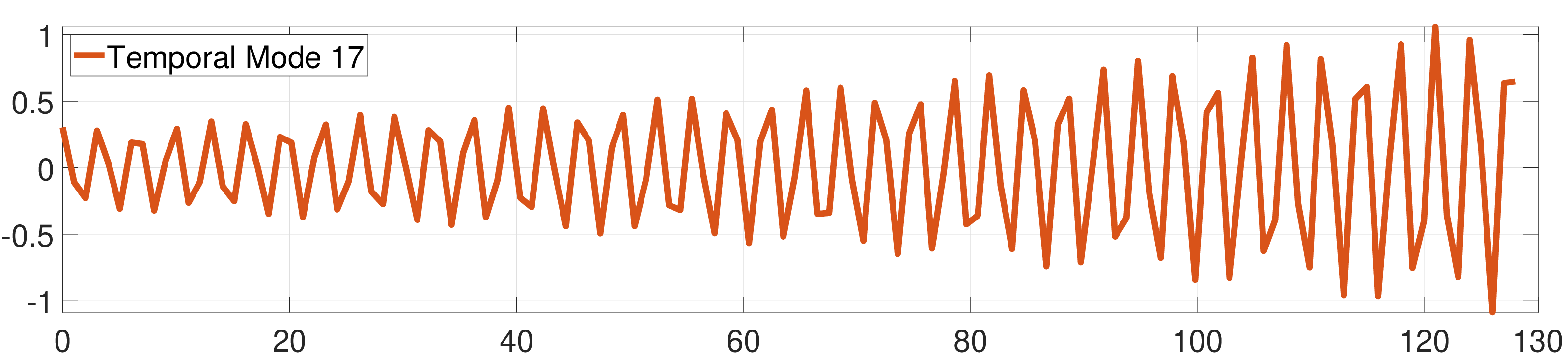}\\
      \includegraphics[width=0.99\linewidth]{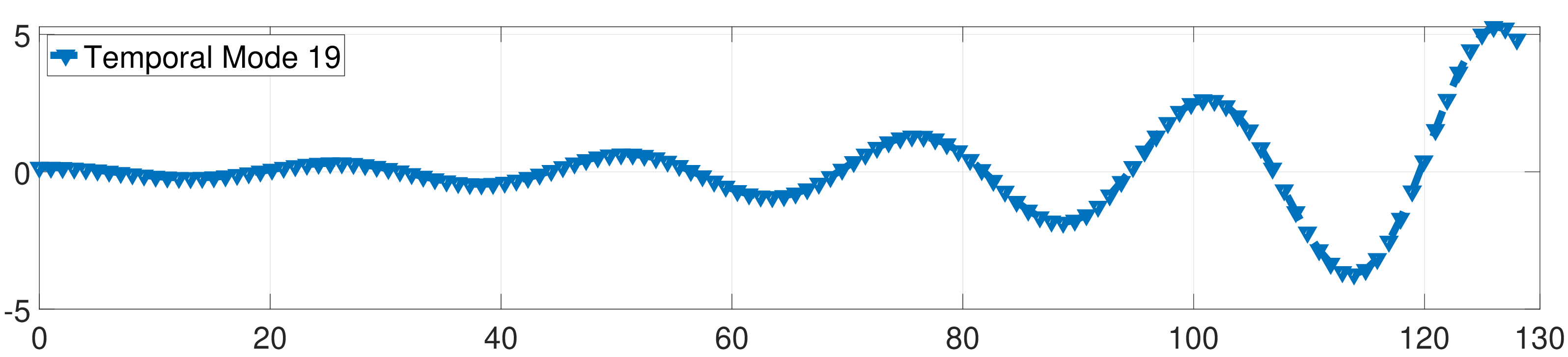}
    \end{subfigure}
         \begin{subfigure}{0.33\textwidth}
     \centering
    \includegraphics[width=0.93\linewidth]{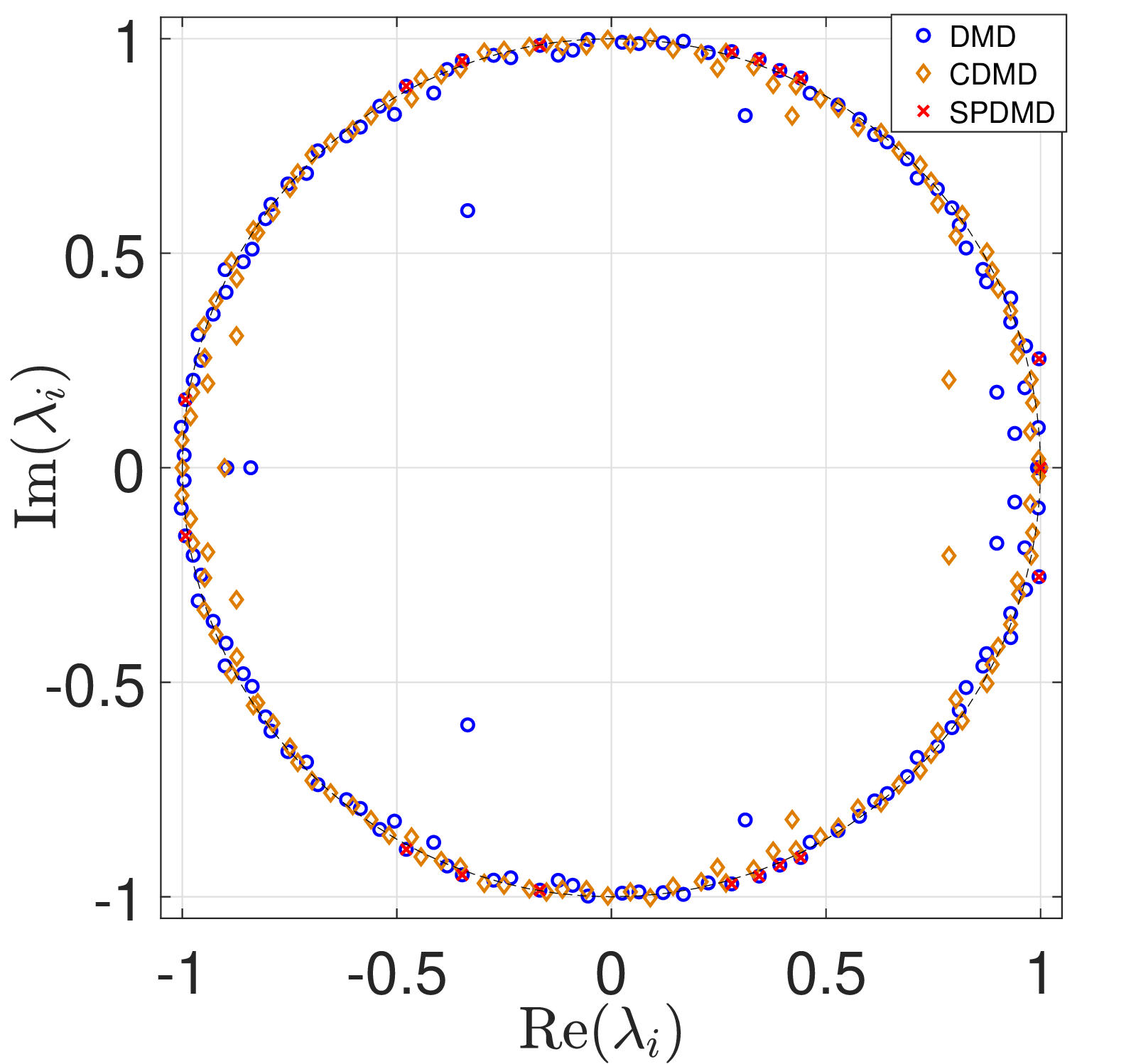} 
    \end{subfigure}
            \begin{subfigure}{0.32\textwidth}
        \centering
     \includegraphics[width=\linewidth]{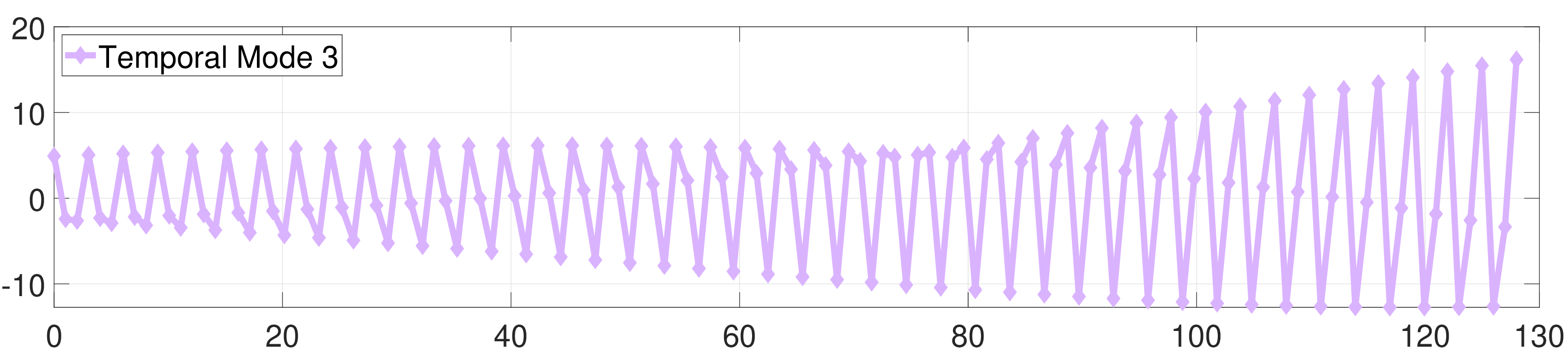}\\
      \includegraphics[width=\linewidth]{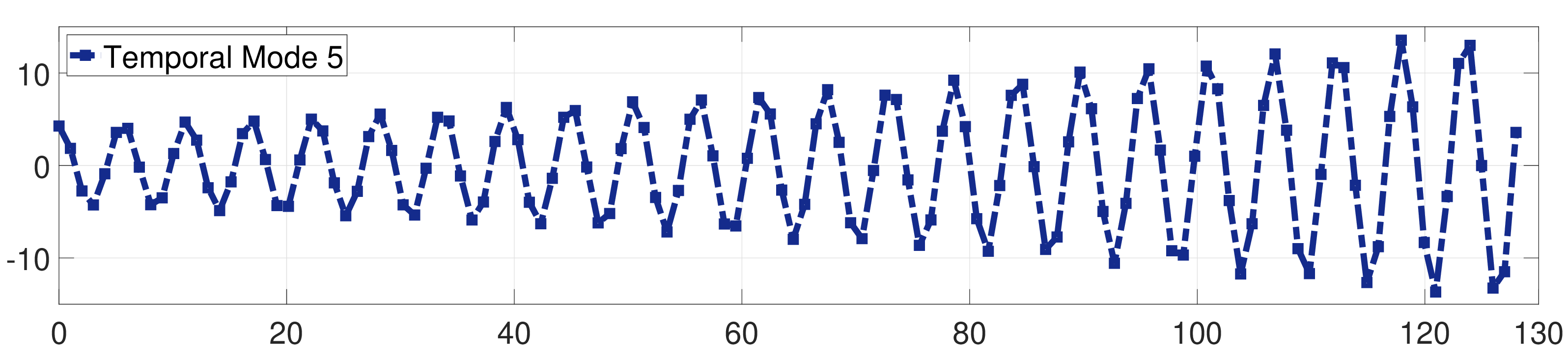}\\
      \includegraphics[width=\linewidth]{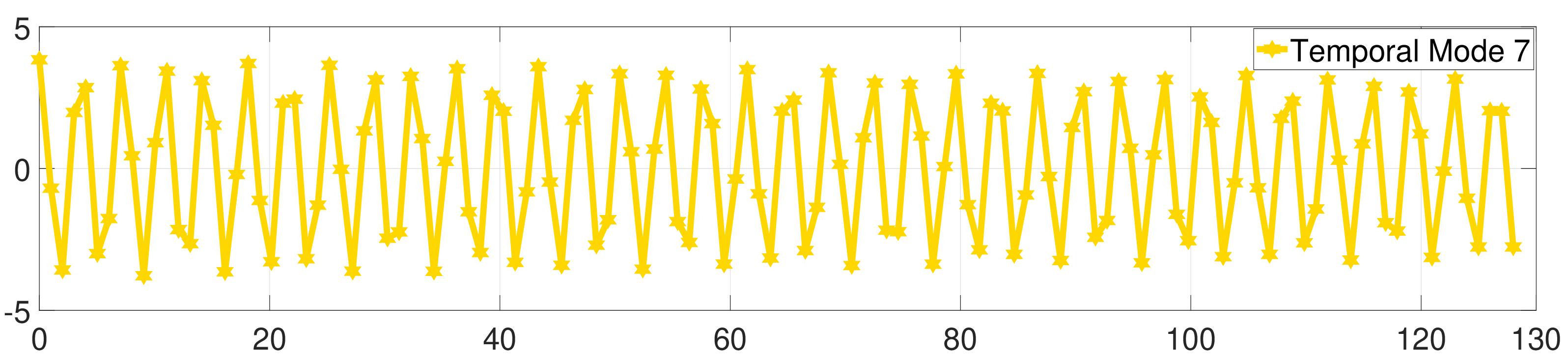}\\
      \includegraphics[width=\linewidth]{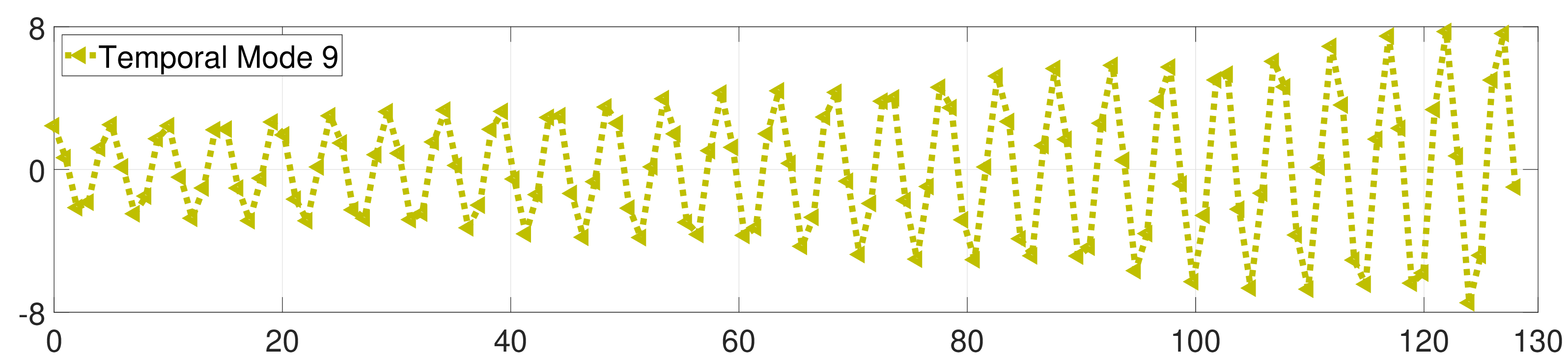}\\
    \end{subfigure}
    \caption{Spectrum distributions of approximated Koopman matrix $A$ w.r.t. annual cycle of SST dataset through DMD ({\color{blue}$\circ$}), CDMD ({\color{orange}$\lozenge$}) and SPDMD ({\color{red}$\times$}), respectively. Here the first mode is ignored since it takes ``Inf'' with nearly zero imaginary part (non-oscillation) with eigenvalues $\lambda_{1}\approx1.000+0000\rm{i}$. The rest plots show the extracted modes via SPDMD.}
    \label{fig:year-eigenvals}
\end{figure*}

By virtue of DMD and SPDMD, Fig.~\ref{fig:Koopman-eigvals-quarterly-SST-case} (left and mid-I) showed the distribution of Koopman eigenvalues on seasonal cycle of (or quarterly) SST data case, exhibiting a distinct ``ring'' structure. 
This sharp contrasts with the monthly SST case shown on the left of Fig.~\ref{fig:Koopman-Eigvals-distributions}, where most eigenvalues lie within the unit circle, while the ring sharp is also similar to that observed in the CDMD in Subsection~\ref{subsec:experiment-MSST}. 
Fig.~\ref{fig:Koopman-eigvals-quarterly-SST-case} (mid-II) plots the growth or decay rate $\mathrm{Re}(\log\lambda_i)$ against the frequency $\mathrm{Im}(\log\lambda_i)$ for the $19$ modes. Furthermore,
Fig.~\ref{fig:Koopman-eigvals-quarterly-SST-case} (right) depicts the trade-off between the accuracy and model-reduction of modes, governed by the range of $\gamma$ values, with $\gamma_{\min} = 0.0001$ and $\gamma_{\max} = 16000$, using $500$ grids, where the function $\bfb(\gamma)$ reduces the number of modes from $511$ to only $4$, and the performance loss function $\Pi\%$ varies from $0.6\%$ to $3.58\%$.
%we refer to see Table~\ref{tabel:SST-quarter} for more details.
On the other hand, the sparsity level of the amplitudes achieved by the solution $\bfb^{\star}$ depends on the weight parameter $\gamma\geq{0}$. Hence, $\bfb(\gamma)$ quantifies the sparsity of the solution, giving a measure of whether a new $\gamma$ results in a sparser solution or not.
The interplay between $\bfb$ and $\gamma$ is formalized through the concept of appropriateness: we say that the weight $\gamma$ is appropriate for the idea amplitude $\bfb$ if a user-specified satisfaction condition is fulfilled, ensuring that the corresponding loss $\Pi\%$ remains acceptable level even with a small number of the captured dominant modes.

SPDMD is applied to determine the number of Koopman modes, where the extracted Koopman modes, in terms of their spatial structure, are shown in Fig.~\ref{fig:spatial-modes-quarter}. The identified $9$ leading modes in the seasonal cycle of the SST data are visualized using the magnitude of the real part of Koopman modes $|{\bm\phi}_{i}|$ for $i=3,5,\ldots,19$.
Meanwhile, Fig.~\ref{fig:quarter-SST-temporal-dynamics} illustrates the real part of the temporal dynamics over seasonal time units. 
% The Koopman eigenfunctions remain on invariant level sets, indicating persistent periodic behavior.
With an eye toward model reduction, 
SPDMD captures the dominant modes associated with those (quasi) periodic behavior (e.g., see Table~\ref{tabel:mode-norm-eigvals-periodic-quarter}) and can also be used for model reduction near periodic Koopman eigenfunctions, simplifying the system's behavior from a large number of degrees of freedom to a low-dimensional representation while preserving its essential dynamics \cite{mezic2005spectral}.

%----------------------loss ===========
\begin{figure}[h!]
    \centering
    \includegraphics[width=0.49\linewidth]{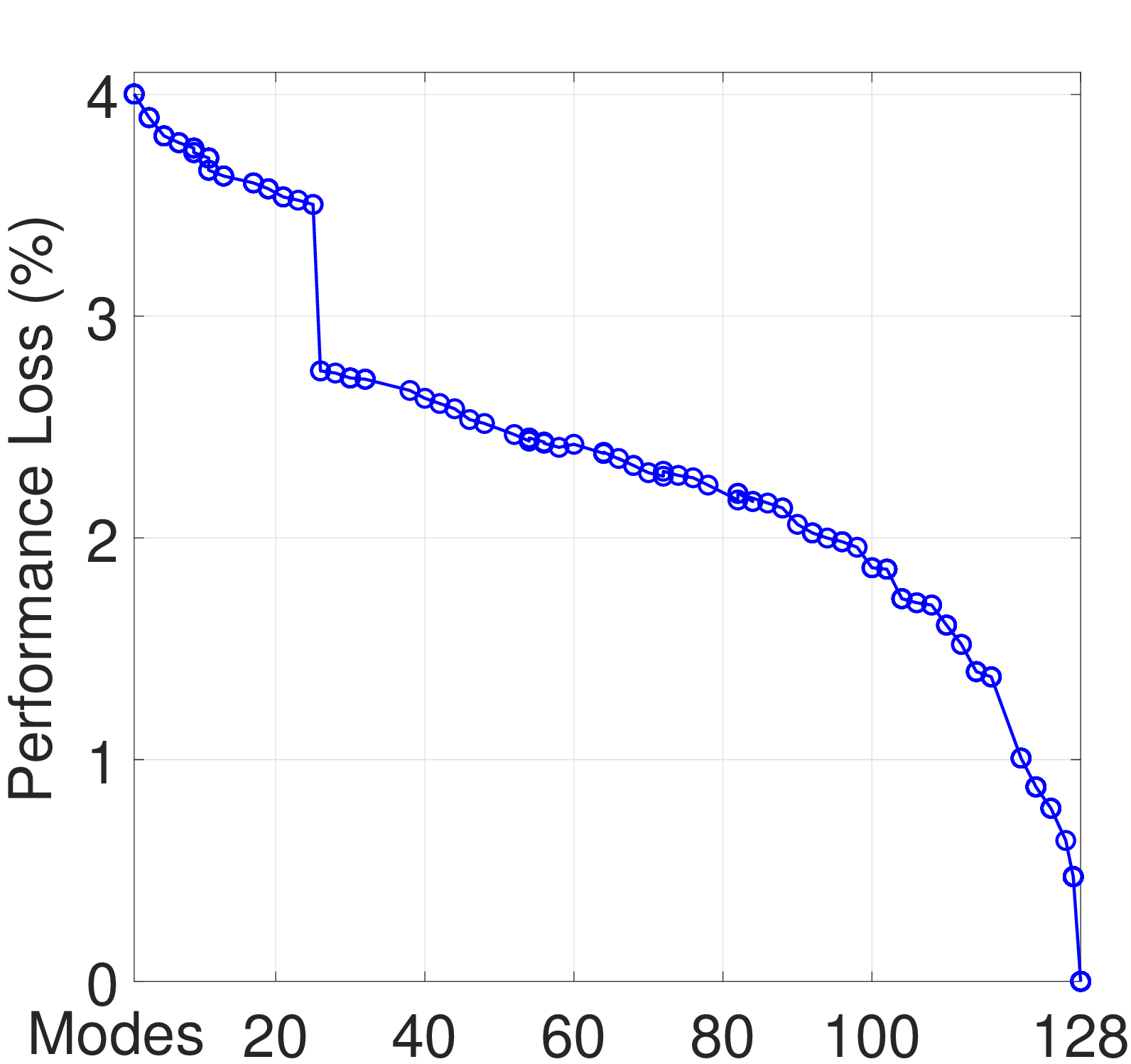}
    \includegraphics[width=0.49\linewidth]{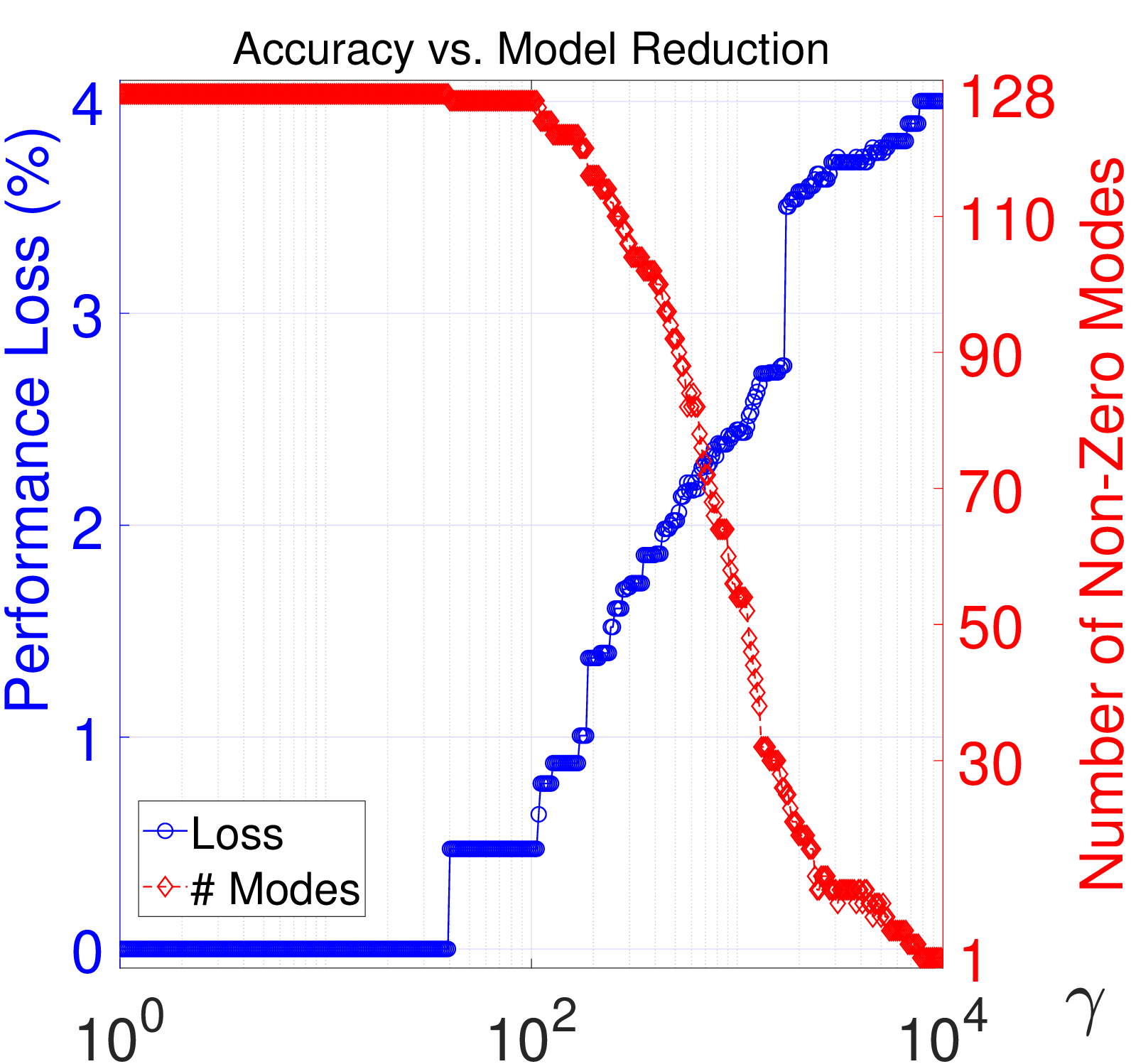}
    \caption{Loss vs. nonzero modes in annual cycle of SST.}
    % # 323 to # 327
    % J_loss $4.653\times{10}^{5}$ to $7.6832\times{10}^{5}$ 
    \label{fig:year-loss}
\end{figure}

\subsection{Experiment~3: Annual Cycle of SST Data Field}\label{subsec:experiment-YSST}
\begin{figure*}[h!]
    \centering
    \includegraphics[width=0.95\linewidth]{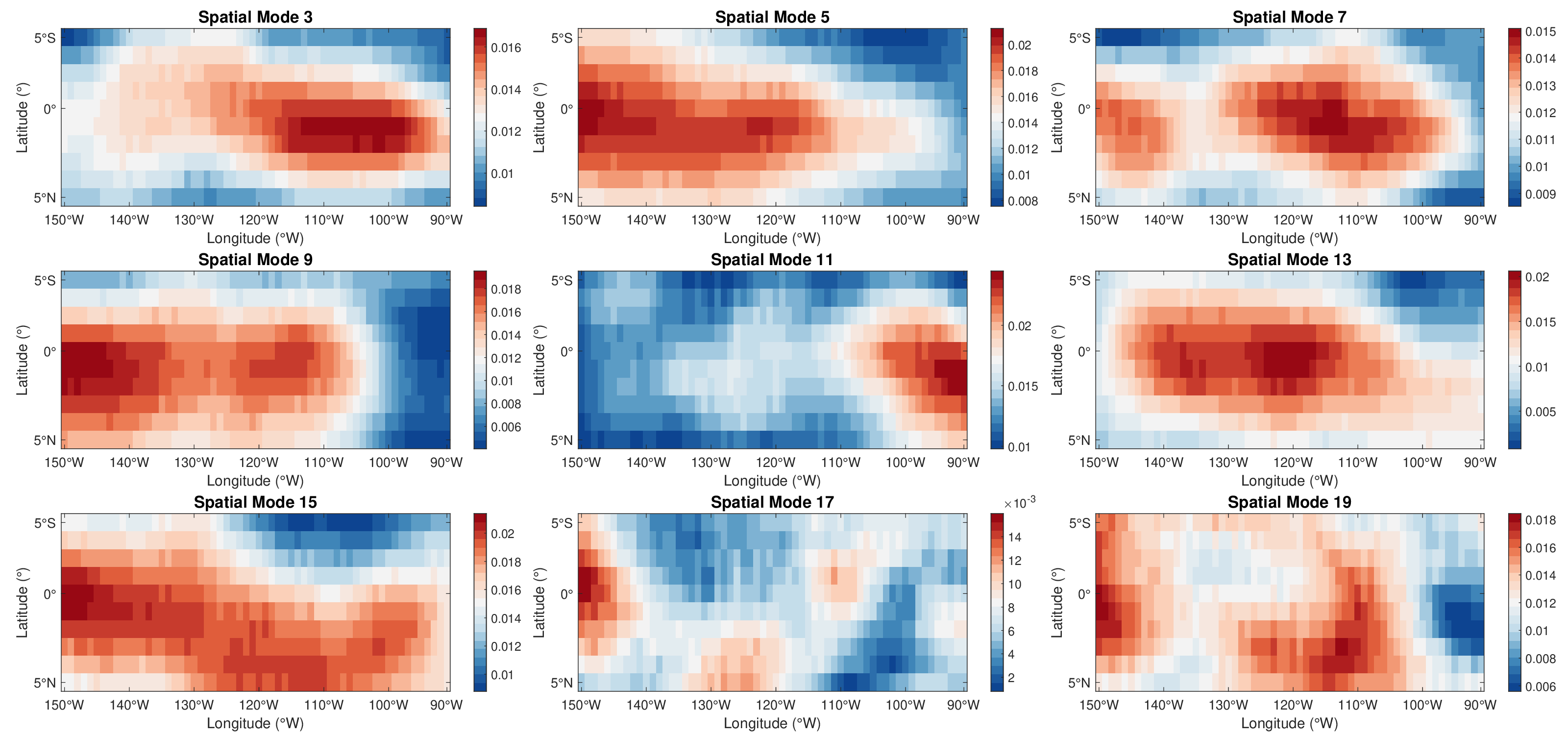}
    \caption{The magnitude of the spatial modes for annual cycle of SST.}
    \label{fig:year-spatial-modes-abs}
\end{figure*}

Finally, we study the experiment on the annual cycle of SST data field. The analysis focuses on short-term climate dynamics with yearly snapshots $N=128$ and low resolution (i.e., a data point per year with $p=7200$), leading to a data matrix $\bfY\in{\mathbb{R}}^{7200\times{128}}$.
Similar to the trials above, the distributions of the Koopman eigenvalues obtained using DMD, CDMD, and SPDMD are labeled with different markers and displayed in Fig.~\ref{fig:year-eigenvals}. Clearly, SPDMD effectively captures the dominant modes with large norm of amplitudes. Along with the information on eigenvalues and amplitudes, Fig.~\ref{fig:year-eigenvals} gives the time evolution of temporal dynamics from the observed annually SST data. Fig.~\ref{fig:year-loss} provides the trade-off between the least-squares approximation error and the number of reduced modes, where the sparsity weight varies as $\gamma\in[1,10^4]$ across $400$ grid points. Obviously, a dramatic jump happens in the number of modes, decreasing from $26$ to $21$, while the loss increases from $2.7533$ to $3.5379$. This transition corresponds to a change in the sparsity weight from $1.6907\times{10}^{3}$ to $1.8543\times{10}^{3}$. 
By selecting a suitable value of $\gamma$, Fig.~\ref{fig:year-spatial-modes-abs} shows the element-wise of absolute value of the spatial modes on 
annual cycle case of SST data.

%=========================spatimal modes on year=====================

%  gamma_grd = 400; % Number of gammaval values
% min_gamma = 1; % Minimum gammaval  
%  max_gamma =1e4; % Maximum gammaval
% % Generate gammaval values using logspace
%  gammaval = logspace(log10(min_gamma), log10(max_gamma), gamma_grd);

%\subsection{Another Look at Continuous and Pseudo Spectra}
%\section{Discussions: Continuous and Pseudo Spectra}\label{sec:continuous-spectrum}

\section{Conclusions}\label{sec:conclusion}
%\subsection{Discussion: continuous spectra}

This paper extracted the dominant spatial and temporal modes of climate dynamics from SST data fields using KMD and SPDMD. 
The leading modes captured reveal the variability of climate dynamics across different cycles (e.g., monthly, seasonal, and annual) of the SST data. 
By tuning the sparsity weights, we achieved a balance between accuracy and model complexity. 
Future work will focus on exploring climate modes through data assimilation or ensemble methods, incorporating a broader range of SST fields and land data.

%========acknowledgement
%\section*{Acknowledgment}
%{\bf Acknowledgment}: 
%We thank Prof. Atsushi Okazaki for his guidance in using the data. 
%This work was partially supported by JST Moonshot R\&D, Grant Number JPMJMS224.

%%   Appendix for Koopman operator theory 

\section*{APPENDIX}\label{appendix}
\subsection{Koopman Operator on Dynamical Systems}\label{appendix:Koopman-operator}
Consider a continuous-time dynamical
system
\begin{align}
 \dot{\bfx}&=\bfF(\bfx), \quad \bfx\in{\mathcal{M}}\subseteq{\mathbb{R}}^{n}
  \label{eq:nonlinear-system} \\
  y&= g(\bfx),
  \label{eq:nonlinear-system-output}
\end{align}
where $\bfx(0)=\bfx_{0}$ is the initial condition, $\bfx$ denotes the state evolving on a finite-dimensional state-space $\mathcal{M}$, 
and the vector field $\bfF:\mathcal{M}\to\mathcal{M}$ 
% $\bfF:\mathcal{M}\to\mathbb{T}_{\bfx}\mathcal{M}$ (tangent space) 
is a possibly smooth map
% from $\mathcal{M}$ ($\mathbb{T}_{\bfx}\mathcal{M}$) to its tangent space $\mathbb{T}_{\bfx}\mathcal{M}$ 
  that describes the nonlinear dynamics over the time $t\geq{0}$. Given the continuous differentiable \emph{scalar} function $g:\mathcal{M}\to{\mathbb{R}}$ %(or ${\mathbb{C}}$) 
which acts on the observable space $\mathcal{F}$, it represents the \emph{output function} of the state associated with the system \eqref{eq:nonlinear-system}. 
The function $g$ defined in \eqref{eq:nonlinear-system-output}, also known as \emph{observable}, can be directly collected from measured data, like historical time series data in the geophysical systems. 
Thus, the nonlinear dynamics $\bfF(\cdot)$ can be treated as \emph{model-free} since true dynamics, such as those in climate systems, often go beyond mathematical toy models (e.g., Lorenz'96 or chaotic models).
We now define a \emph{flow map} ${S}^{t}:\mathcal{M}\to\mathcal{M}$ by taking the trajectories of \eqref{eq:nonlinear-system}, which is a one parameter semigroup of diffeomorphsims $\{S^{t}\}_{t\geq{0}}$ \cite[Chap.~1]{mauroy2020koopman}.

%\begin{definition}\label{def:Koopman-operator}
Alternatively, Koopman \cite{koopman1931hamiltonian} proved that nonlinear systems \eqref{eq:nonlinear-system} 
can be recast in terms of Koopman operator ${\mathcal{K}}^{t}:{\mathcal{F}}\to{\mathcal{F}}$ associated with the flow $S^{t}$ and observables $g$ in \eqref{eq:nonlinear-system-output}, defined by
\begin{align}
{\mathcal{K}}^{t}g=g\circ{S^{t}},\quad \forall{g}\in{\mathcal{F}},
    \label{eq:Koopman-operator}
\end{align}
where the observable space ${\mathcal{F}}$ is possibly an {infinite-dimensional} (Banach) space.
The Koopman operator is a \emph{bounded, linear} operator that describes the {linear} evolution of observables and provides a global perspective on the dynamics and moves our attention from the state-space variables to the observable functions\footnote{Koopman dynamics is dictated by the infinitesimal generator ${\mathcal{L}}_{g}$, 
%of Koopman operator, 
i.e., $\dot{g}
\!=\!\lim_{t\downarrow{0}}({{\mathcal{K}}^{t}g\!-\!g})/{t}
\!=\!({\bfF}\cdot\nabla){g}
\!:=\!{\mathcal{L}}_{g}$, which is expressed by the inner product of the nonlinear field and the gradient $\nabla$ of the observables based on the chain rule and the limit exists in the strong sense \cite{mauroy2020koopman}.}.%e.g., see Fig.~\ref{fig:Koopman-framework}. 
%(or the functions of the state variables). %in terms of the linear combination.

Substituting Koopman operator \eqref{eq:Koopman-operator} into the output \eqref{eq:nonlinear-system-output} along the observables, we have the decomposition:
\begin{align}
    y(t)=g(\bfx(t))=g(S^{t}(\bfx_{0}))=({\mathcal{K}}^{t}g)(\bfx_{0}).
    \label{eq:Koopman-decomp-output-1}
\end{align}

  An eigenfunction of Koopman operator is an observable ${\varphi_{j}\in{\mathcal{F}}\setminus\{0\}}$, associated with eigenvalue $\nu_{j}\in{\mathbb{C}}$, satisfying
  %if for some eigenvalue $\lambda_{j}$ it satisfies
  \begin{align}
  {\mathcal{K}}^{t}\varphi_{i}=\varphi_{i}(S^{t}(\bfx_{0}))=e^{\nu_{i}{t}}\varphi_{j},
  \quad j=1,2,\ldots
      \label{eq:KEFS-KEs}
  \end{align}
% for some eigenvalue $\lambda\in{\mathbb{C}}$. 
where $\varphi_{j}$ is the Koopman eigenfunction (KEF), and  $\nu_{j}\in{\mathbb{C}}$ is named as Koopman eigenvalue (KE) that belongs to \emph{point} or \emph{discrete spectrum} of the bounded linear operator $\mathcal{K}$ to mainly capture the dynamics information in terms of the periodic steady-state, transient (damped-oscillatory mode), and regular pattern \cite{mauroy2020koopman}.

\subsection{Koopman Mode Decomposition}
Let $\bfx=[x_{0},\ldots,x_{n}]$ be a point in the state space $\mathcal{M}$, the discrete-time dynamics is characterized as follows\footnote{The discrete-time case can be derived from the continuous-time nonlinear system \eqref{eq:nonlinear-system}, using a sampling period $h:=t_{k+1}-t_{k}$.% For details on the Koopman operator, refer to Appendix \ref{appendix:Koopman-operator}.
}
%and a time-discretization of \eqref{eq:nonlinear-system} is employed a sampling period $h:=t_{k+1}-t_{k}$ such that 
\begin{align}
   \bfx_{k+1}=\bfT(\bfx_{k}),\quad k=0,1,\ldots,N-1.
    \label{eq:DT-systems}
\end{align}
where $\bfx_{k}=\bfx(kh)=\bfx(t_{k})$ and the discrete-time map $S^{t}_{h}(\cdot):=\bfT(\cdot)$ evolves $\bfx_{k}$ to $\bfx_{k+1}$ \cite{susuki2011nonlinear}. Analogous expression holds for discrete-time Koopman eigenvalue (KE) and Koopman eigenfunction (KEF), \eqref{eq:KEFS-KEs} replaced by ${\mathcal{K}}{\varphi}_{j}={\lambda}_{j}{\varphi}_{j}$ if we assume ${\lambda}_{j}=e^{\nu_{j}{t}}$. Moreover,
the vector-valued observable $\bfg:{\mathcal{M}}\to{\mathbb{R}}^{p}$ defined by 
 \begin{align}
   \bfg(\bfx)=[g_{1}(\bfx),\cdots,g_{p}(\bfx)]^{\top}, 
   \label{eq:vector-valued-observable}
 \end{align}
can be expanded in terms of diverse combinations of the eigenfunctions $\{\varphi_{j}(\bfx)\}_{j=1}^{\infty}$. 
In this context, the vector-valued \emph{full-state} observables $\bfy=\bfg(\bfx)=\bfx$ when $p=n$, simply returns the state vector itself with the same dimensionality.
For example,
if all the elements of observable $\bfg$ 
in \eqref{eq:vector-valued-observable} 
lie within the span of the eigenfunctions $\varphi_{j}$, 
then it can be derived in terms of a (countable) set of these KEFs $\bfg(\bfx)=\sum_{j=1}^{\infty}\varphi_{j}(\bfx)\bfv_{j}$,
where $\bfv_{j}\in{\mathbb{C}}^{p}$ is the well-known \emph{Koopman modes} (KM) as a vector-valued coefficient for the expansion.
Besides, Koopman operator $\mathcal{K}$ evolves the observable %\eqref{eq:Koopman-decomp-output-2} 
according to \eqref{eq:Koopman-decomp-output-1}, yielding the Koopman mode decomposition (KMD) \cite{mezic2005spectral,mauroy2020koopman} as follows 
\begin{align}
    {\mathcal{K}}^{t}(\bfg(\bfx_{0}))\!=\!{\mathcal{K}}^{t}\left(\sum_{j=1}^{\infty}\varphi_{j}(\bfx_{0})\bfv_{j}\right)
    \!=\!\sum_{j=1}^{\infty}\lambda_{j}^{t}\varphi_{j}(\bfx_{0})\bfv_{j},
    \label{eq:Koopman-decomp-output-3}
\end{align}
where the elements $\{\lambda_{j},\varphi_{j},\bfv_{j}\}$, $j=1,2,\ldots,$ or $\infty$ are referred to as \emph{Koopman tuples}. The Koopman eigenfunctions and Koopman eigenvalues encode lots of information about the underlying dynamical systems.
Finite-dimensional approximations of the Koopman operator $\mathcal{K}$ rely on a \emph{finite} number of tuples as defined in \eqref{eq:Koopman-decomp-output-3}, which can be computed using projection-based, data-driven DMD methods (see \eqref{eq:DMD-approx} and \cite{tu2014dynamic,kutz2016dynamic} for more details).

\bibliographystyle{IEEEtran}
\bibliography{main}

\end{document}